\documentclass[12pt,noshowpacs,nofootinbib,notitlepage,amsmath]{revtex4-1}
\pdfoutput=1
\allowdisplaybreaks
\usepackage{graphicx,color}
\usepackage[colorlinks=true,citecolor=blue,linkcolor=blue,urlcolor=blue]{hyperref}
\usepackage[charter]{mathdesign}
\DeclareSymbolFontAlphabet{\mathcal}{symbols}
\DeclareSymbolFont{symbols}{OMS}{xmdcmsy}{m}{n}
\DeclareSymbolFont{largesymbols}{OMX}{xmdcmex}{m}{n}
\SetSymbolFont{symbols}{bold}{OMS}{xmdcmsy}{b}{n}

\begin{document}  
\title{\color{blue}\Large Stable Asymptotically Free Extensions\\(SAFEs) of the Standard Model}
\author{Bob Holdom}
\email{bob.holdom@utoronto.ca}
\author{Jing Ren}
\email{jren@physics.utoronto.ca}
\author{Chen Zhang}
\email{czhang@physics.utoronto.ca}
\affiliation{Department of Physics, University of Toronto, Toronto, Ontario, Canada  M5S1A7}
\begin{abstract}
We consider possible extensions of the standard model that are not only completely asymptotically free, but are such that the UV fixed point is completely UV attractive. All couplings flow towards a set of fixed ratios in the UV. Motivated by low scale unification, semi-simple gauge groups with elementary scalars in various representations are explored. The simplest model is a version of the Pati-Salam model. The Higgs boson is truly elementary but dynamical symmetry breaking from strong interactions may be needed at the unification scale. A hierarchy problem, much reduced from grand unified theories, is still in need of a solution.
\end{abstract}
\maketitle

\section{Introduction }
\label{intro}

We start by considering an elementary Higgs boson in a world without low energy supersymmetry. In this world there are two conflicting demands on the nature of new physics on higher mass scales. Naturalness strongly constrains the new physics to prevent unwanted contributions to the Higgs mass. Either the new physics mass scale cannot be much higher than the Higgs mass or the Higgs coupling to the new physics must be extremely weak. The other demand on the new physics is that it must significantly alter the running of couplings, including the quartic coupling of the Higgs. This is because the Landau poles in the quartic coupling and the $U(1)$ hypercharge coupling would signal new mass scales of the dangerous type. To avoid this requires new massive degrees of freedom that do couple to standard model fields and thus are also dangerous for naturalness. These two demands are suggesting that if there is new physics to cure the Landau problem then it must enter at as low a scale as possible to minimize the naturalness problem.

The absence of Landau poles is a requirement for the theory to be UV complete, or in other words that there is a description of the theory on arbitrarily high energy scales in terms of elementary fields. The fermions and gauge bosons of asymptotically free gauge theories are prime examples of truly elementary fields. The standard model is not of this type, but it often thought that there is no reason it should be given the presence of gravity. The onset of gravitational effects at Planckian energies is usually taken to mean that the theory experiences a complete change of character on these scales. But once again this is at odds with naturalness. It is only if gravity somehow exerts only a very minimal effect on the scalar sector in a UV complete theory is there is any hope of naturalness.

There have been recent attempts to show how the effects of gravity in UV complete quantum field theories could be consistent with naturalness. Ref.~\cite{Dubovsky:2013ira} illustrated a proposed mechanism in a 2D model of quantum gravity. These authors introduce the concept of ``gravitational dressing'' of a QFT, where Planck mass effects modify the S-matrix directly without inducing any physical mass scales. Ref.~\cite{Salvio:2014soa} (see also \cite{Einhorn:2014gfa}) suggests that the pure gravitational action in the high energy regime just contains two terms, an $R^2$ term and the Weyl term $\frac{1}{3}R^2-R_{\mu\nu}^2$. The Einstein-Hilbert term is induced via the VEV of a new scalar field with non-minimal coupling to $R$.  The point is that the gravitational interactions may then be both renormalizable and asymptotically free \cite{Stelle:1976gc}\cite{Julve:1978xn}\cite{Fradkin:1981iu}
. Ref.~\cite{Salvio:2014soa} argues that such a gravity sector could be arranged to couple sufficiently weakly to the standard model fields to preserve naturalness. The gravity sector here is not quite complete because of a ghost and a tachyon in the spectrum.

Our interest here is the other half of the problem, how to build UV complete quantum field theories containing truly elementary scalar fields. We approach this by searching for gauge theories containing both fermions and scalars where all couplings run to zero in the UV. This could provide a completely asymptotically free extension (CAFE) of the standard model. A nice study of this type was conducted long ago in \cite{Cheng:1973nv}. There the constraints were found on theories with a simple gauge group with varying numbers of scalar fields in various representations and with fermions. Gauge, quartic and Yukawa couplings were considered. CAFEs were found and described in terms of UV fixed points (UVFPs) where ratios of couplings approached fixed values. The fixed points were also required to be UV attractive from all directions in coupling space. Thus these are CAFEs that also have complete UV stability, and we denote such an extension of the standard model as a SAFE. That such theories were found in \cite{Cheng:1973nv} may have been of interest to the construction of grand unified theories. But the study showed that it was difficult for the scalars that were allowed to sufficiently break down the original gauge theory via the Higgs mechanism. For this reason and perhaps also because it was thought that gravity would nevertheless provide an ultraviolet cutoff, it appears that SAFEs were never considered to be of particular importance in GUTs.

Our work can be considered to be a continuation of this old work. Since we need to embed the standard model into a gauge group without a $U(1)$ factor at the lowest possible scale we are here dealing with low scale unification. Thus we must extend the original work to semi-simple gauge groups. A minimal requirement is that the scalar content of the theory must yield the Higgs doublet after symmetry breaking. We don't require that the scalars be entirely responsible for gauge symmetry breaking, other than electroweak symmetry breaking, since we leave open the possibility that strong interactions could dynamically break some symmetries.

After the work \cite{Cheng:1973nv} there were attempts to find other realistic CAFEs, not necessarily grand unified. From our point of view these attempts were not completely successful since UV stability was dropped (see review \cite{Callaway:1988ya} and references therein and in particular \cite{Kalashnikov:1976hr}). The fixed point was allowed to be UV repulsive in some directions in coupling space. In this case the space of couplings that do flow to the fixed point has reduced dimensionality. This amounts to constraints (sometimes called predictions) on the low energy couplings that are also affected by higher order corrections. Satisfying the constraints would require fine tuning the couplings order by order in perturbation theory. In our work we shall insist on complete UV stability.

Much more recently there has been another attempt to find UV complete theories with elementary scalars, but this time the search was for \emph{nontrivial} UVFPs \cite{Litim:2014uca}. Unlike the case of asymptotic freedom, here the fixed point requires knowledge of the $\beta$-functions beyond lowest order. Interesting examples were found but here again complete UV stability was not attained. Also, in this context the work in \cite{Tavares:2013dga} suggests that the transition from a regime of running couplings to a nontrivial UVFP is sufficient to cause a contribution to the Higgs mass. So in this case as well, the corresponding mass scale must be as low as possible.

The prototype of low scale unification is the Pati-Salam model \cite{Pati:1974yy}, based on the gauge group $SU(4)\times SU(2)_L\times SU(2)_R$, with the fermions of one family in the (4, 2, 1)$_L +($4, 1, 2)$_R$ representation. Our study will answer the question as to whether scalars can be added such that a SAFE results. But we shall set up our study in a more general context where we consider products of various $SU(N)$ gauge groups with various scalars that may transform simultaneously under two or three of these gauge groups. We only consider scalars in the fundamental representation since then we can expect a Higgs doublet to emerge after symmetry breaking. These results may be of more general interest for model building.

Since we are discussing theories that are UV complete above the Planck scale, one might wonder about the effect of gravity on the running couplings of the matter fields. This was discussed in the quadratic higher derivative gravity theories of \cite{Salvio:2014soa,Einhorn:2014gfa}. The coupling $f_2^2$, appearing as $1/f_2^2$ times the Weyl term, is always asymptotically free with both gravity and matter fields contributing with the same sign to the $\beta$-function. This means that $f_2^2$ is typically much smaller than the gauge couplings in the deep UV, and so its effect can be neglected. The coupling $f_0^2$ appearing in the $R^2$ term will be asymptoticallly free only if the ratio $f_0^2/f_2^2$ becomes negative in the UV. Depending on the matter content it is possible that $f_0^2$ could run relatively slowly and thus play a more significant role. Here we note a discrepancy in the calculated $f_0^2$ contribution to the scalar quartic $\beta$-functions in \cite{Salvio:2014soa} and \cite{Einhorn:2014gfa}. In the following we shall ignore the possible effect of gravity on the matter $\beta$-functions.

This paper is organized as follows. In Sec.~\ref{review} we first review the basic idea to realize SAFEs with a simple Lie group. Then we generalize the study to a semi-simple gauge group in Sec.~\ref{general}, as motivated by low scale unification. For quantitative study we choose several benchmarks for gauge groups and scalar representations. In Sec.~\ref{results} we present and discuss the numerical results. Based on these studies we consider the simplest example of a SAFE with low scale unification in Sec.~\ref{models}. We conclude in Sec.~\ref{conclusion}.

\section{SAFE\lowercase{s} with Simple Lie Group}
\label{review}

In this section we review the basic idea to realize SAFEs in \cite{Cheng:1973nv}. This reference systematically studied the simple group $SU(N)$ or $O(N)$ case with fermions and scalars in various representations. Here we supplement their work with some numerical results for comparison with our later analysis.

Since we study UV asymptotic freedom, the one loop $\beta$-functions are sufficient to study the UV behavior. At one loop the coupled $\beta$-functions of gauge, Yukawa and quartic couplings can be solved sequentially. For the gauge coupling, its $\beta$-function only depends on itself and yields
\begin{align}\label{eq:gaugebeta}
\beta=\frac{dg}{dt}=\frac{bg^3}{(4\pi)^2}\Rightarrow
g^2(t)=-\frac{8\pi^2}{bt}
\end{align}
with $t=\ln(\mu/\Lambda)$. $b<0$ gives asymptotic freedom with an infrared Landau pole at $t=0$ ($\mu=\Lambda$). The $\beta$-function coefficient $b$ gives the running speed of gauge coupling at large $t$. For the Yukawa coupling $y$, its $\beta$-function has the generic form
\begin{align}
(4\pi)^2\beta_y=a_yy^3-a_gg^2y,
\end{align}
where $a_y, a_g>0$. The dependence on $g$ can be eliminated with a change of variables $\bar{y}\equiv y^2/g^2$, and this gives
\begin{align}\label{eq:YukawaBeta2}
(4\pi)^2g^{-2}\beta_{\bar{y}}=2\bar{y}\left[a_y\bar{y}-(a_g+b)\right]
\end{align}
where dependence on $b$ has appeared. To have asymptotically free $y$ amounts to finding a UVFP for $\bar{y}$. When $a_g+b\leq0$ and since $\bar{y}\geq0$ by definition the only UVFP is $\bar{y}=0$, which is UV repulsive. A stable UVFP requires that $a_g+b>0$ in which case $\bar{y}=0$ is the stable UVFP. The result is that $\bar y$ decreases asymptotically as
\begin{align}
\bar y(t)\propto t^\frac{a_g+b}{b}.
\end{align}
 As clarified in \cite{Cheng:1973nv}, the same conclusion applies to the more complicated case when the Yukawa couplings are described by a matrix. So in SAFEs, the contribution of Yukawa couplings is negligible in the $\beta$-functions of quartic couplings in the deep UV.

The one loop $\beta$-function of a scalar quartic coupling is in general a function of both gauge and Yukawa couplings, but as we have just explained we ignore the latter. We may illustrate the general features with one scalar $\Phi_i$ in the fundamental representation of a $SU(N_A)$ gauge group. The gauge invariant scalar potential at $\text{dim}=4$ has only one term,
\begin{align}\label{eq:potentialsNA}
  V_4=\lambda\,\Phi^{*}_{i}\Phi^{}_{i}\Phi^{*}_{j}\Phi^{}_{j}
.\end{align}
The one loop $\beta$-function for $\lambda$ is
\begin{align}\label{eq:betaNA}
  (4\pi )^2\beta _{\lambda}=4\left(N+4\right)\lambda^2
  -6\lambda g^2\left(N-\frac{1}{N}\right)
  +\frac{3(N-1)(N^2+2N-2)}{4N^2}g^4.
\end{align}
This $\beta$-function is composed of three pieces: the positive pure quartic terms, negative gauge-quartic terms and positive pure gauge terms. To have $\beta_\lambda=0$ the three contributions should be comparable and so this disfavors a large hierarchy between quartic and gauge couplings. In particular quartic couplings must also run as $1/t$ in the deep UV.

We may again eliminate the dependence on $g$ by a change of variables $\bar\lambda\equiv\lambda/g^2$, and this gives
\begin{align}\label{eq:betaNAbar}
  (4\pi )^2g^{-2}\beta _{\bar\lambda}=4\left(N+4\right)\bar\lambda^2
  -\bar\lambda\left[2b+6\left(N-\frac{1}{N}\right)\right]
  +\frac{3(N-1)(N^2+2N-2)}{4N^2}
\end{align}
with $b$ again appearing in the linear term. Defining $r\equiv b/b_M$ where $b_M=-11N/3$ is the pure gauge boson contribution, the regions with $2b+6(N-\frac{1}{N})>0$ and $2b+6(N-\frac{1}{N})<0$ meet at the value $r_0=\frac{9}{11}(1-1/N^2)$. These two regions correspond to the slow gauge running $(r_s<r_0)$ and fast gauge running ($r_f>r_0$) cases respectively, and there is a one-to-one mapping with $2r_sb_{M}+6(N-\frac{1}{N})=-(2r_fb_{M}+6(N-\frac{1}{N}))$ and $\lambda\to-\lambda$.

$\beta_{\bar\lambda}=0$ is simply a quadratic equation for $\bar\lambda$ and there are two real roots when
\begin{align}\label{eq:NAbAcond}
  \left[2b+6\left(N-\frac{1}{N}\right)\right]^2-\frac{12}{N^2}(N+4)(N-1)(N^2+2N-2)>0.
\end{align}
This inequality sets an upper (lower) bound on $r$ in the slow (fast) running region with solutions $\bar\lambda>0$ ($\bar\lambda<0$). For the present example the lower bound on $r$ in the branch $r>r_0$ is always above one and so this cannot be realized with any matter assignment. Also this region is disfavored due to the upper bound on $r$ from the UV stability of Yukawa coupling and the vacuum stability condition for the quartic coupling $\bar\lambda>0$. So we need only consider the slow running region, where the inequality (\ref{eq:NAbAcond}) sets $N\geq3$.

\linespread{1}\begin{figure}[!h]
  \centering%
{ \includegraphics[width=10cm]{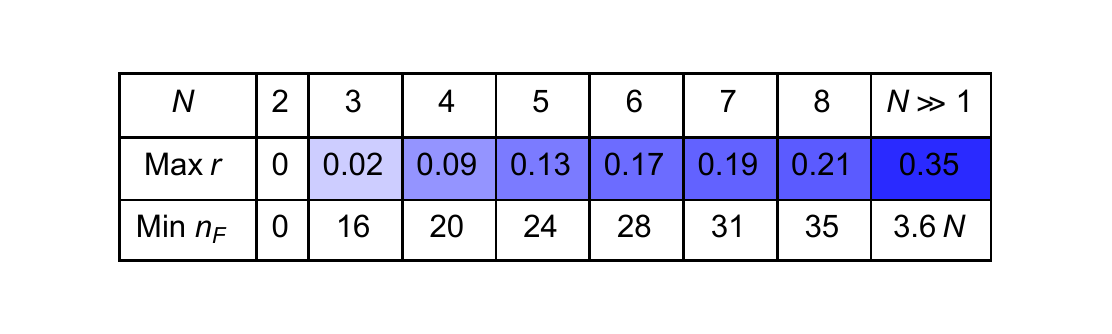}}
\caption{\label{fig:NAmatrix} The maximum $r=b/b_M$ for one fundamental scalar of $SU(N)$ and the minimum number $n_F$ of Dirac fundamental fermions to achieve this.}
\end{figure}

For each $N\geq3$, we present the upper bound on $0\leq r\leq1$ for various $N$ in Fig.~\ref{fig:NAmatrix}. We can determine the number of Dirac fermions $n_F$ to satisfy this bound from
\begin{align}
b=b_M+n_F b_F+\frac{1}{6}.
\end{align}
The minimum $n_F$ basically grows with $N$, and it is shown for the fundamental representation $b_F=2/3$ in the last row in Fig.~\ref{fig:NAmatrix}.

\linespread{1}\begin{figure}[!h]
  \centering%
{ \includegraphics[width=7cm]{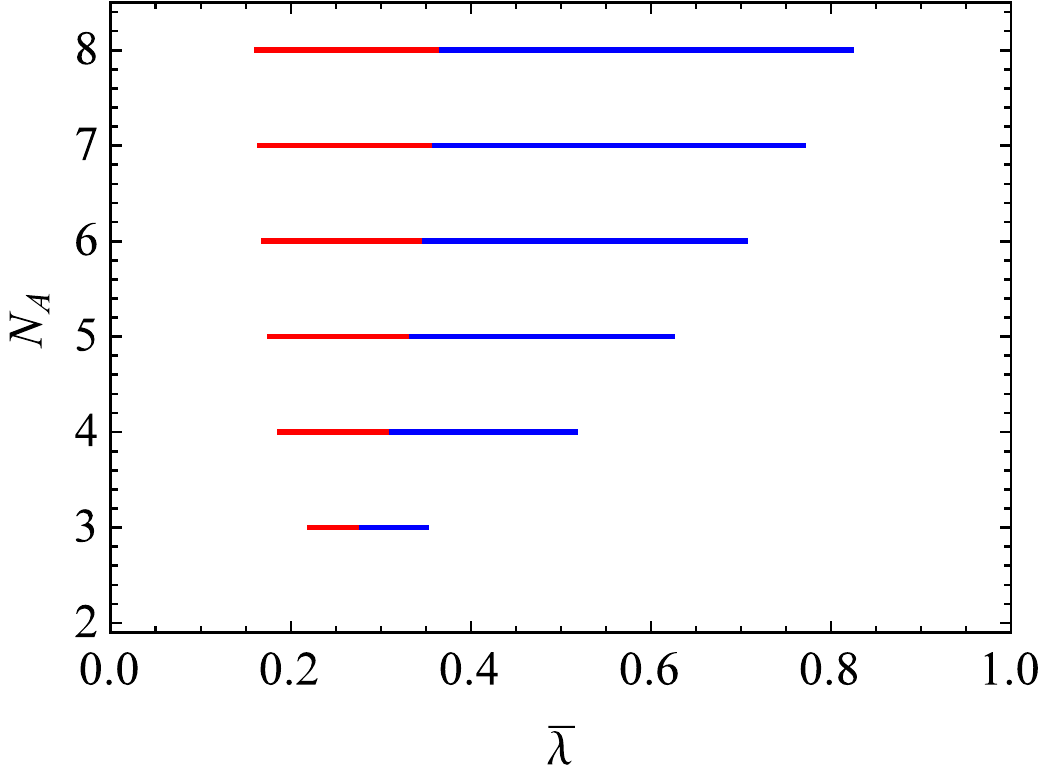}}
\caption{\label{fig:lambdaNA} The values of $\bar\lambda=\lambda/g^2$ at stable (red) and unstable (blue) UVFPs as $r$ varies over the allowed range.}
\end{figure}

For each $N, b$ that satisfy (\ref{eq:NAbAcond}) and $r<r_0$  there are two positive real roots $\bar\lambda_1<\bar\lambda_2$. Given the positive contribution from the pure quartic and pure gauge terms, it is the smaller root $\bar\lambda_1$ that is stable, i.e. $d\beta/d\bar\lambda<0$ at $\bar\lambda=\bar\lambda_1$. For each $N$ we depict $\bar\lambda_1, \bar\lambda_2$ for all possible $b$ in Fig.~\ref{fig:lambdaNA}, where red and blue label stable and unstable UVFPs respectively. $b\to0$ at the ends of each line. In large $N\gg1$ limit, the stable and unstable UVFPs become insensitive to $N$ and these end values approach $0.14$ and $1.3$ respectively. For a stable UVFP, $\bar\lambda$ is always smaller than one. Also, the stable UVFP $\bar\lambda_1$ is UV attractive with respect to all quartic couplings $\bar\lambda<\bar\lambda_2$.

By increasing the size and/or number of scalar representations, a larger $N$ may be required to achieve a SAFE. This generally does not allow sufficient scalar fields to break the simple gauge group in some realistic manner \cite{Cheng:1973nv}. For example $SU(5)$ grand unification typically requires two scalars, in the adjoint and fundamental representations, to break $SU(5)$ down to the SM. But with this set of scalars the theory is a SAFE only if $N\geq7$.

For a given gauge group, the larger the total number of scalar degrees of freedom, the tighter is the constraint on $b$ \cite{Cheng:1973nv}. This general feature will carry over to our generalizations and it is another motivation to restrict ourselves to scalars in the fundamental representation.

\section{Generalization to Semi-simple Lie Group}
\label{general}

Motivated by low scale unification we shall focus on scalar fields transforming under the following two types of gauge groups with $N_i\geq 2$.
\begin{align}\label{eq:semiLie}
(1):\,SU(N_A)\times SU(N_B),\quad
(2):\,SU(N_A)\times SU(N_B)\times SU(N_C)
\end{align}
We first discuss the behavior of Yukawa couplings for the semi-simple case. In the simplest case of a single Yukawa coupling $y$, as a generalization of the $\beta$-function in (\ref{eq:YukawaBeta2}) we find
\begin{align}\label{eq:YukawaBeta3}
(4\pi)^2g_j^{-2}\beta_{\bar{y}}=2\bar{y}\left[a_y\bar{y}-\sum_{i\neq j}a_i g_i^2/g_j^2-(a_j+b_j)\right]
,\end{align}
where $\bar{y}=y^2/g_j^2$ and with $g_j$ one of the gauge couplings. The $a_i$ depend on the scalar and fermion representations. In the deep UV the gauge coupling $g_i$ approaches its asymptotic form and becomes insensitive to its initial value. So we may replace the ratio of gauge couplings in (\ref{eq:YukawaBeta3}) by their $\beta$-functions coefficients, i.e. $g_i^2/g_j^2\to b_j/b_i$. If
\begin{align}\label{eq:YukawaSUVFPcond}
1+\sum_{i}\frac{a_i}{b_i}<0
\end{align}
then there is a stable UVFP and it is at $\bar{y}=0$.

We have checked various fermion and scalar representations for the gauge groups in (\ref{eq:semiLie}). It turns out that (\ref{eq:YukawaSUVFPcond}) is easy to satisfy since $a_i\sim N_i$ and $b_i$ is negative. In some cases (\ref{eq:YukawaSUVFPcond}) may put a upper bound on $b_i$, but as we shall see below, in the parameter space of interest the constraint is much weaker than constraints from the quartic couplings. For a matrix of Yukawa couplings we expect these features will continue to hold, as in \cite{Cheng:1973nv}. Therefore in our study of SAFEs for semi-simple gauge group we will focus on the quartic couplings and neglect the contribution of Yukawa couplings in their $\beta$-functions.

We now build four benchmarks for semi-simple Lie groups in (\ref{eq:semiLie}).

\subsection*{Case A: $SU(N_A)\times SU(N_B)$ with $(N_A, N_B)$}
For the gauge group $SU(N_A)\times SU(N_B)$ the simplest nontrivial setup is to have one scalar field $\Phi_{ik}$ that transforms in the fundamental representation of both groups, i.e. $(N_A, N_B)$. The most general $\text{dim}=4$ scalar potential is
\begin{align}\label{eq:potentialsNANB}
V_4=\lambda_{d}\Phi^{*}_{ik}\Phi^{}_{ik}\Phi^{*}_{jl}\Phi^{}_{jl}+\lambda_{s}\Phi^{*}_{ik}\Phi^{}_{il}\Phi^{*}_{jl}\Phi^{}_{jk}
\end{align}
 when at least one $N_i>2$. $\lambda_d$ and $\lambda_s$ denote double trace and single trace couplings respectively. In the deep UV, the $\beta$-functions for these quartic couplings are
\begin{align}\label{eq:betaNANB}
(4\pi )^2\beta _{\lambda _{d}}&=4\left[\left(N_AN_B+4\right)\lambda _{d}^2+2\left(N_A+N_B\right)\lambda _{d}\lambda _{s}+3\lambda _{s}^2\right]
-6\lambda _{d}\left[\left(N_A-\frac{1}{N_A}\right)g_A^2\right.\nonumber\\
&\left.+\left(N_B-\frac{1}{N_B}\right)g_B^2\right]
+\frac{3}{4}\left[\left(1+\frac{2}{N_A^2}\right)g_A^4+\left(1+\frac{2}{N_B^2}\right)g_B^4\right]+3g_A^2g_B^2\left(1+\frac{1}{N_AN_B}\right)\nonumber\\
(4\pi )^2\beta _{\lambda _{s}}&=4\lambda _{s}\left[(N_A+N_B)\lambda _{s}+6\lambda _{d}\right]
-6\lambda _{s}\left[\left(N_A-\frac{1}{N_A}\right)g_A^2+\left(N_B-\frac{1}{N_B}\right)g_B^2\right]\nonumber\\
&+\frac{3}{4}\left[\left(N_A-\frac{4}{N_A}\right)g_A^4+\left(N_B-\frac{4}{N_B}\right)g_B^4\right]-3g_A^2g_B^2\left(\frac{1}{N_A}+\frac{1}{N_B}\right)
\end{align}
It is straightforward to verify that (\ref{eq:betaNANB}) reduces to (\ref{eq:betaNA}) in the single gauge group case with $N_B\to1, g_B\to0$ and $\lambda_d+\lambda_s\to \lambda$. The $N_A=N_B=2$ case corresponds to the bidoublet in the left-right symmetric model and it has a larger set of couplings \cite{LRM}.

\subsection*{Case B: $SU(N_A)\times SU(N_B)$ with $(N_A, N_B)$ and $(N_A,1)$}
In the second benchmark we consider the same gauge group with two scalars. We don't expect to learn much by considering two copies of $(N_A, N_B)$, especially since the replication of scalars was considered in \cite{Cheng:1973nv}. For the combination $(N_A,1)+(1,N_B)$ there is a limit where the two scalars decouple and so this case is also of not much interest. So we will study two different scalars that share a common gauge group.
\begin{align}\label{eq:scalar2}
\Phi^{(1)}_{ik}: (N_A,\,N_B),\quad
\Phi^{(2)}_{j}: (N_A,\,1)
\end{align}
$N_A$ specifies the common gauge group. The most general scalar potential when at least one $N_i>2$ has five terms,
\begin{align}\label{eq:potential2}
V_4&=\lambda_{d1}\Phi^{(1)*}_{ik}\Phi^{(1)}_{ik}\Phi^{(1)*}_{jl}\Phi^{(1)}_{jl}+\lambda_{s1}\Phi^{(1)*}_{ik}\Phi^{(1)}_{il}\Phi^{(1)*}_{jl}\Phi^{(1)}_{jk}
+\lambda_{2}\Phi^{(2)*}_{i}\Phi^{(2)}_{i}\Phi^{(2)*}_{j}\Phi^{(2)}_{j}\nonumber\\
&+2\lambda_{d12}\Phi^{(1)*}_{ik}\Phi^{(1)}_{ik}\Phi^{(2)*}_{j}\Phi^{(2)}_{j}+2\lambda_{s12}\Phi^{(1)*}_{ik}\Phi^{(1)}_{jk}\Phi^{(2)*}_{j}\Phi^{(2)}_{i}
.\end{align}
Here there are two mixing couplings $\lambda_{d12}, \lambda_{s12}$. The one loop $\beta$-functions are presented in (\ref{eq:beta2}) in Appendix A. Due to the presence of the common gauge group we shall find that there is no UVFP solution where the mixing couplings vanish and the two scalars decouple.

\subsection*{Case C: $SU(N_A)\times SU(N_B)\times SU(N_C)$ with $(N_A, N_B, 1)$ and $(N_A, 1, N_C)$}

With the enlarged gauge symmetry $SU(N_A)\times SU(N_B)\times SU(N_C)$, the next interesting scalar content starts with two scalars. It is again interesting to study the case with two different scalars sharing a common gauge group. The case different from Case B is the following.
\begin{align}\label{eq:scalar3}
\Phi^{(1)}_{ik}: (N_A,\,N_B,\,1),\quad
\Phi^{(2)}_{ja}: (N_A,\,1,\,N_C)
\end{align}
We set $N_A>2$ for the common gauge group. In the context of the Pati-Salam model, this setup may correspond to left-right symmetric scalars $(4,2,1)$ and $(4,1,2)$. The scalar potential is
\begin{align}\label{eq:potential3}
V_4&=\lambda_{d1}\Phi^{(1)*}_{ik}\Phi^{(1)}_{ik}\Phi^{(1)*}_{jl}\Phi^{(1)}_{jl}+\lambda_{s1}\Phi^{(1)*}_{ik}\Phi^{(1)}_{il}\Phi^{(1)*}_{jl}\Phi^{(1)}_{jk}
+\lambda_{d2}\Phi^{(2)*}_{ia}\Phi^{(2)}_{ia}\Phi^{(2)*}_{jb}\Phi^{(2)}_{jb}+\lambda_{s2}\Phi^{(2)*}_{ia}\Phi^{(2)}_{ib}\Phi^{(2)*}_{jb}\Phi^{(2)}_{ja}\nonumber\\
&+2\lambda_{d12}\Phi^{(1)*}_{ik}\Phi^{(1)}_{ik}\Phi^{(2)*}_{ja}\Phi^{(2)}_{ja}+2\lambda_{s12}\Phi^{(1)*}_{ik}\Phi^{(1)}_{jk}\Phi^{(2)*}_{ja}\Phi^{(2)}_{ia},
\end{align}
where $\lambda_{d12}, \lambda_{s12}$ are mixing couplings. We may consider a simplified version of this theory by imposing a $Z_2$ symmetry, the analogy of left-right symmetry in the Pati-Salam model.
\begin{align}\label{eq:caseC1Z2}
\textrm{Case C1}\,(Z_2\,\,\textrm{symmetry}):\,\,N_B=N_C,\,\,g_B=g_C,\,\,\lambda_{d2}=\lambda_{d1},\,\,\lambda_{s2}=\lambda_{s1}
\end{align}
This Case C1 amounts to picking a special slice in the whole parameter space, with only two gauge couplings and four quartic couplings. The $\beta$-functions are presented in (\ref{eq:beta31}).

We denote by case C2 the general case with six quartic couplings. The $\beta$-functions are in (\ref{eq:beta32}).

In the case of the Pati-Salam model with $\Phi_L=(4,2,1)$ and $\Phi_R=(4,1,2)$ we may construct a gauge invariant quartic term with the Levi-Civita symbol,
\begin{align}\label{eq:potential31LC}
V_4\supset \frac{1}{2}\lambda_{\epsilon}\epsilon_{iji'j'}\epsilon_{kl}\epsilon_{k'l'}\left[\Phi^{(1)}_{ik}\Phi^{(1)}_{jl}\Phi^{(2)}_{i'k'}\Phi^{(2)}_{j'l'}+h.c.\right].
\end{align}
This amounts to Det($\Phi$) for $4\times4$ matrix $\Phi\equiv(\,\Phi_L\,\,\Phi_R\,)$, which vanishes for $\Phi_L=\Phi_R$. The modified $\beta$-functions with the $\lambda_\epsilon$ contribution are presented in (\ref{eq:beta32LC1}), (\ref{eq:beta32LC2}).

\subsection*{Case D: $SU(N_A)\times SU(N_B)\times SU(N_C)$ with $(N_A, N_B, N_C)$}

In the last benchmark we study a scalar representation charged under all three groups. In particular we consider the fundamental representation $\Phi_{ika}:(N_A, N_B, N_C)$. This type of scalar field is less studied in literature since its VEV breaks all gauge symmetries at the same scale. But in view of finding SAFEs it is intriguing to ask whether it helps to have a scalar transforming under more gauge groups. The scalar potential is
\begin{align}\label{eq:potentialsNANBNC}
V_4&=\lambda_{d}\Phi^{*}_{ika}\Phi^{}_{ika}\Phi^{*}_{jlb}\Phi^{}_{jlb}+\lambda_{s1}\Phi^{*}_{ika}\Phi^{}_{jka}\Phi^{*}_{jlb}\Phi^{}_{ilb}\nonumber\\&+\lambda_{s2}\Phi^{*}_{ika}\Phi^{}_{ila}\Phi^{*}_{jlb}\Phi^{}_{jkb}
+\lambda_{s3}\Phi^{*}_{ika}\Phi^{}_{ikb}\Phi^{*}_{jlb}\Phi^{}_{jla}.
\end{align}
There are now three single trace couplings. The one loop $\beta$-functions are presented in (\ref{eq:betaNANBNC}), and they are symmetric under interchanges between $(N_A,\lambda_{s1})$, $(N_B,\lambda_{s2})$ and $(N_C,\lambda_{s3})$. One can verify that (\ref{eq:betaNANBNC}) reduces to (\ref{eq:betaNANB}) with $N_C\to1$, $g_C\to0$ and $\lambda_d+\lambda_{s3}\to\lambda_d$, $\lambda_{s1}+\lambda_{s2}\to\lambda_s$.

In the Pati-Salam model with one $(4,2,2)$ scalar we may construct another Levi-Civita term,
\begin{align}\label{eq:CaseDLC}
V_4\supset \frac{1}{6}\lambda_{\epsilon}\epsilon_{iji'j'}\epsilon_{kl}\epsilon_{mn}\epsilon_{ac}\epsilon_{bd}\left[\Phi_{ika}\Phi_{jlb}\Phi_{i'mc}\Phi_{j'nd}+h.c.\right].
\end{align}
The $\beta$-functions involving $\lambda_\epsilon$ are presented in (\ref{eq:betaNANBNCLC1}) and (\ref{eq:betaNANBNCLC2}).

\section{Numerical results and analysis}
\label{results}

In this section we present the numerical results and analysis of the four benchmarks. As before we change variables $\bar\lambda_i=\lambda_i/g_j^2$ where $g_j$ is one of the gauge couplings. Then we replace the ratios of different gauge couplings by their  asymptotic values, $g_i^2/g_j^2\to b_j/b_i$. This leaves us with coupled quadratic equations of the $\bar\lambda_i$. Taking case A as an example, the $\beta$-functions in (\ref{eq:betaNANB}) become
\begin{align}\label{eq:betaNANBr}
(4\pi )^2g_A^{-2}\beta _{\bar\lambda _{d}}&=4\left[\left(N_AN_B+4\right)\bar\lambda _{d}^2+2\left(N_A+N_B\right)\bar\lambda _{d}\bar\lambda _{s}+3\bar\lambda _{s}^2\right]
-\bar\lambda _{d}b_A\left[2+\frac{6}{b_A}\left(N_A-\frac{1}{N_A}\right)\right.\nonumber\\
&\left.+\frac{6}{b_B}\left(N_B-\frac{1}{N_B}\right)\right]
+\frac{3}{4}b_A^2\left[\frac{1}{b_A^2}\left(1+\frac{2}{N_A^2}\right)+\frac{1}{b_B^2}\left(1+\frac{2}{N_B^2}\right)+\frac{4}{b_Ab_B}\left(1+\frac{1}{N_AN_B}\right)\right]\nonumber\\
(4\pi )^2g_A^{-2}\beta _{\bar\lambda _{s}}&=4\bar\lambda _{s}\left[(N_A+N_B)\bar\lambda _{s}+6\bar\lambda _{d}\right]
-\bar\lambda _{s}b_A\left[2+\frac{6}{b_A}\left(N_A-\frac{1}{N_A}\right)+\frac{6}{b_B}\left(N_B-\frac{1}{N_B}\right)\right]\nonumber\\
&+\frac{3}{4}b_A^2\left[\frac{1}{b_A^2}\left(N_A-\frac{4}{N_A}\right)+\frac{1}{b_B^2}\left(N_B-\frac{4}{N_B}\right)-\frac{4}{b_Ab_B}\left(\frac{1}{N_A}+\frac{1}{N_B}\right)\right],
\end{align}
where $\bar\lambda_i=\lambda_i/g_A^2$.

With these we can solve for the UVFP of $\{\bar\lambda_i\}$ as functions of $N_i$ and $b_i$. Since the coupled quadratic equations are usually difficult to solve analytically, we find numerical solutions for a parameter scan over $N_i, b_i$. To illustrate the pattern, we choose $2\leq N_i\leq8$. The $\beta$-function coefficients $b_i$ depend on the matter and are model dependent. For convenience we use $r_i\equiv b_i/b_{i,M}$, where $b_{i,M}=-11N_i/3$, and we consider the range $0<r_i\leq 1$. The $\{\bar\lambda_{0,i}\}$ at UVFPs should be real but need not be positive.

To find UV stability we study the RG flows in vicinity of the UVFP. At linear order it is characterized by the matrix
\begin{align}\label{eq:stablematrix}
D_{ij}(\bar\lambda_{0,i})\equiv \left.\frac{\partial \beta_{\bar\lambda_i}}{\partial \bar\lambda_j}\right|_{\bar\lambda_i=\bar\lambda_{0,i}}
.\end{align}
The UVFP is absolutely stable as long as all eigenvalues $\kappa_k$ of $D_{ij}(\bar\lambda_{0,i})$ are negative. The UVFP for the $\bar\lambda_i$'s is approached along the directions of the eigenvectors as $t^{-\kappa_k/2b_A}$.

\subsection{Constraints on $r_i\equiv b_i/b_{i,M}$ from the parameter scan}

\linespread{1}\begin{figure}[!h]
  \centering%
{ \includegraphics[width=4cm]{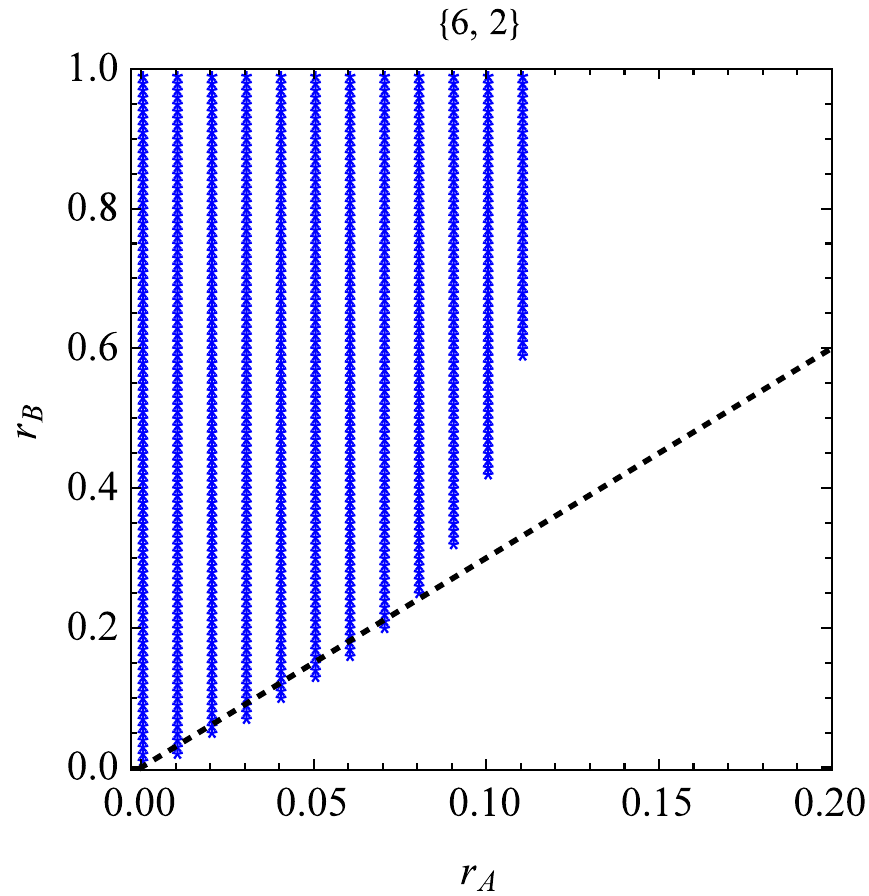}
\includegraphics[width=4cm]{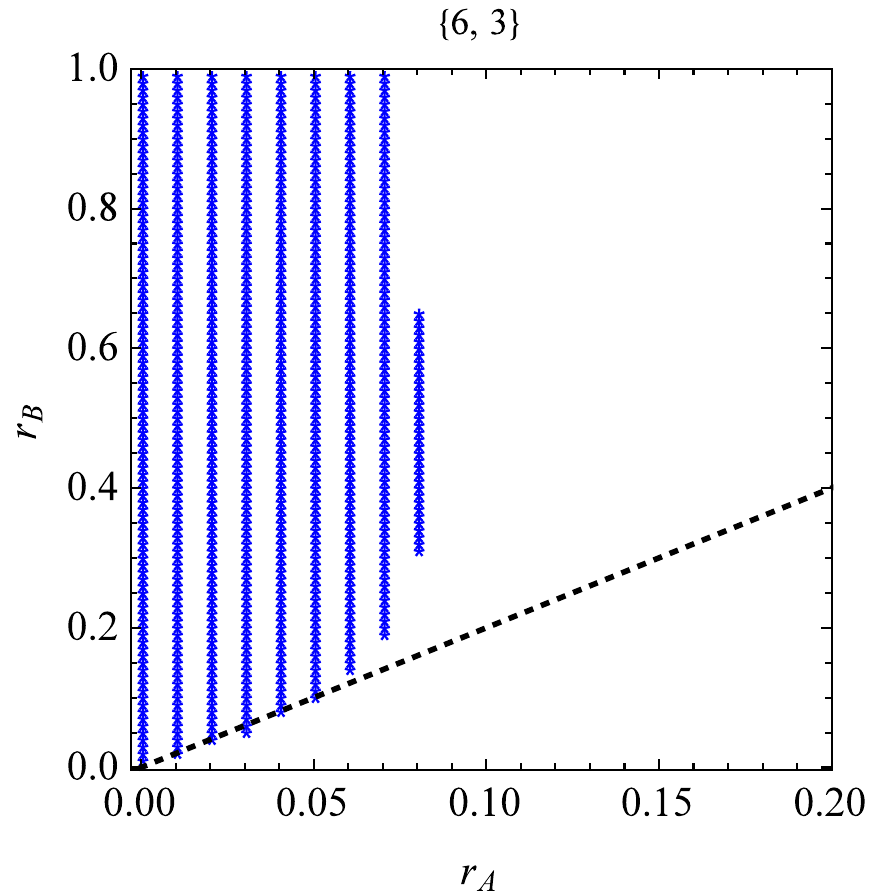}
\includegraphics[width=4cm]{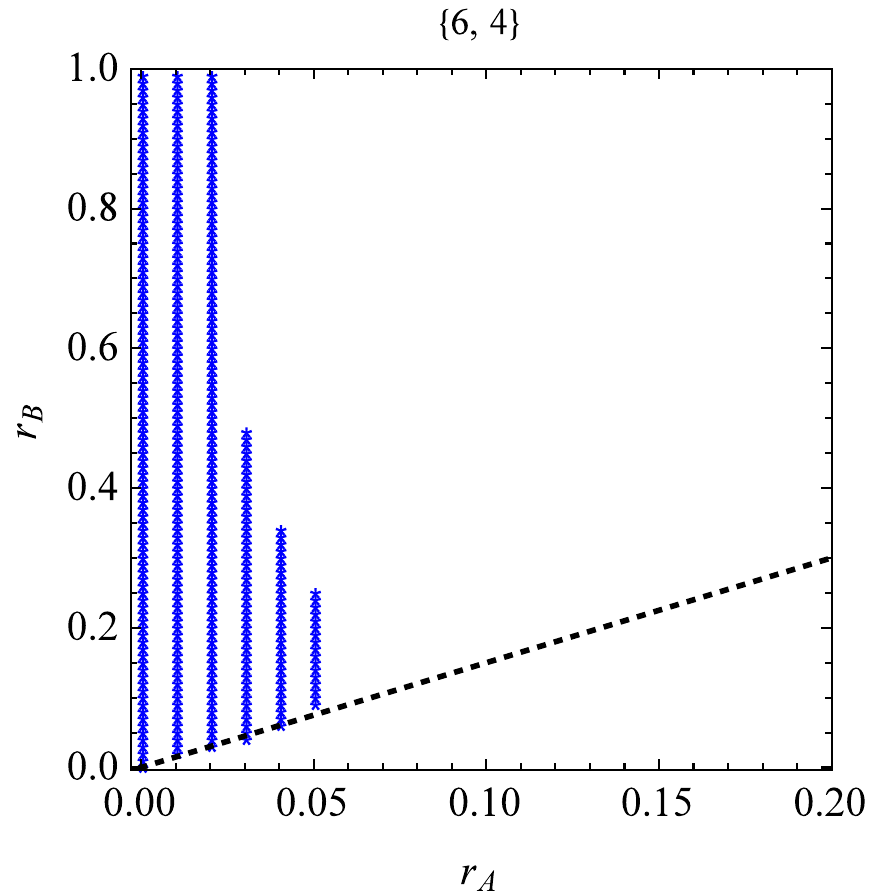}
\includegraphics[width=4cm]{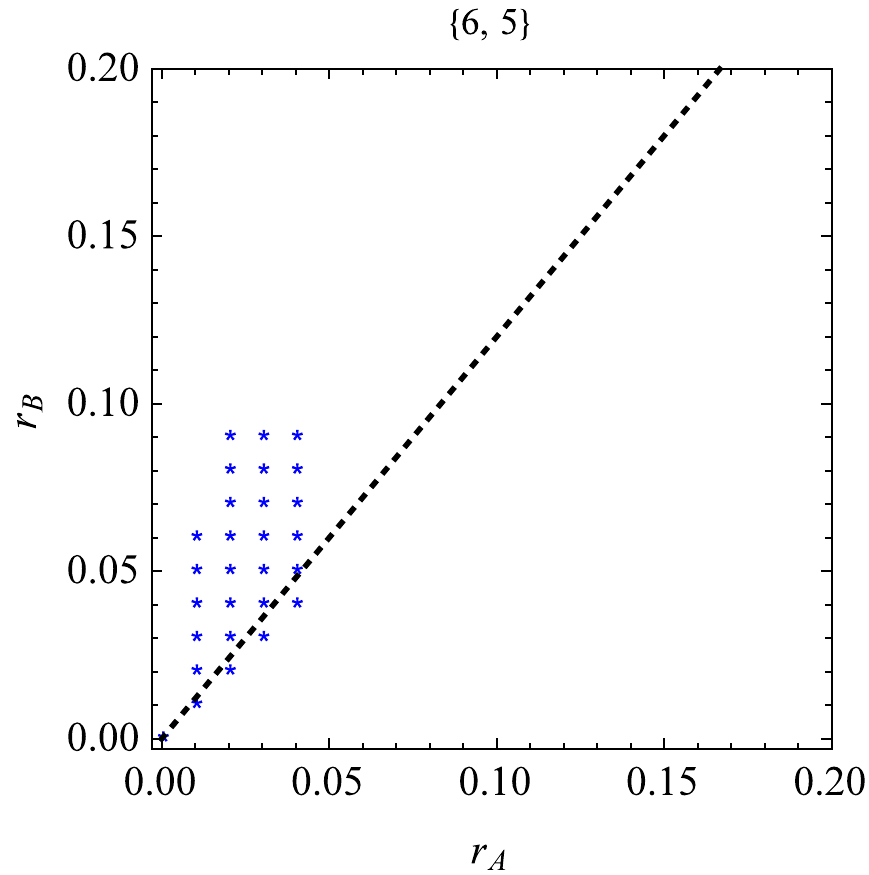}\\
 \includegraphics[width=4cm]{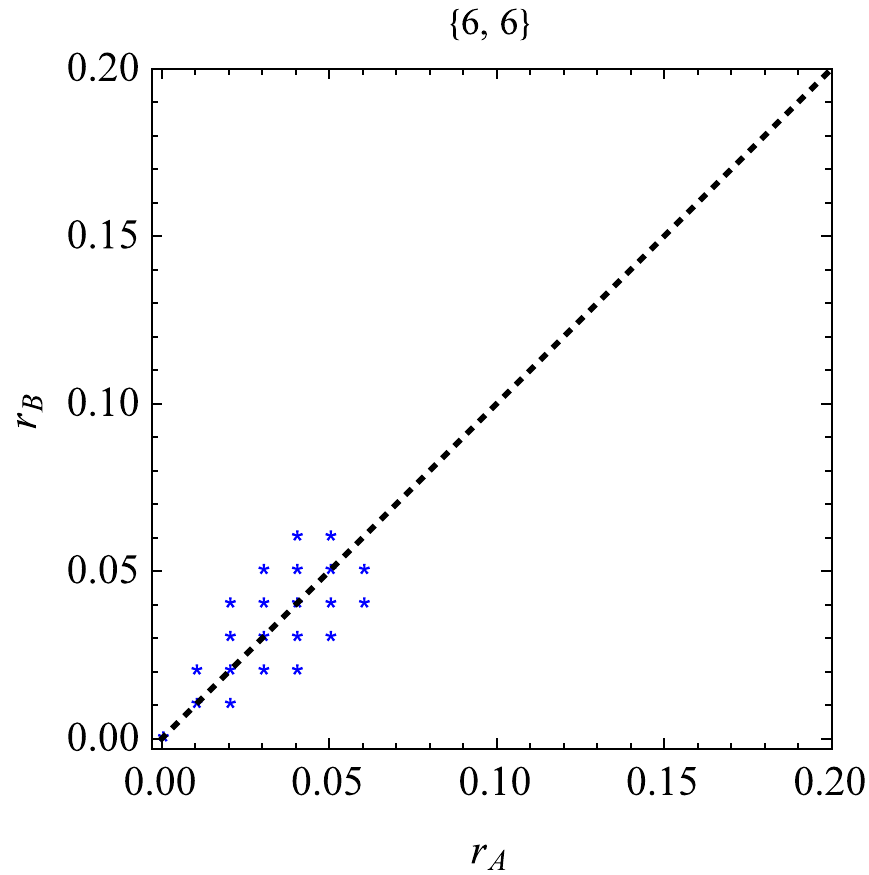}
\includegraphics[width=4cm]{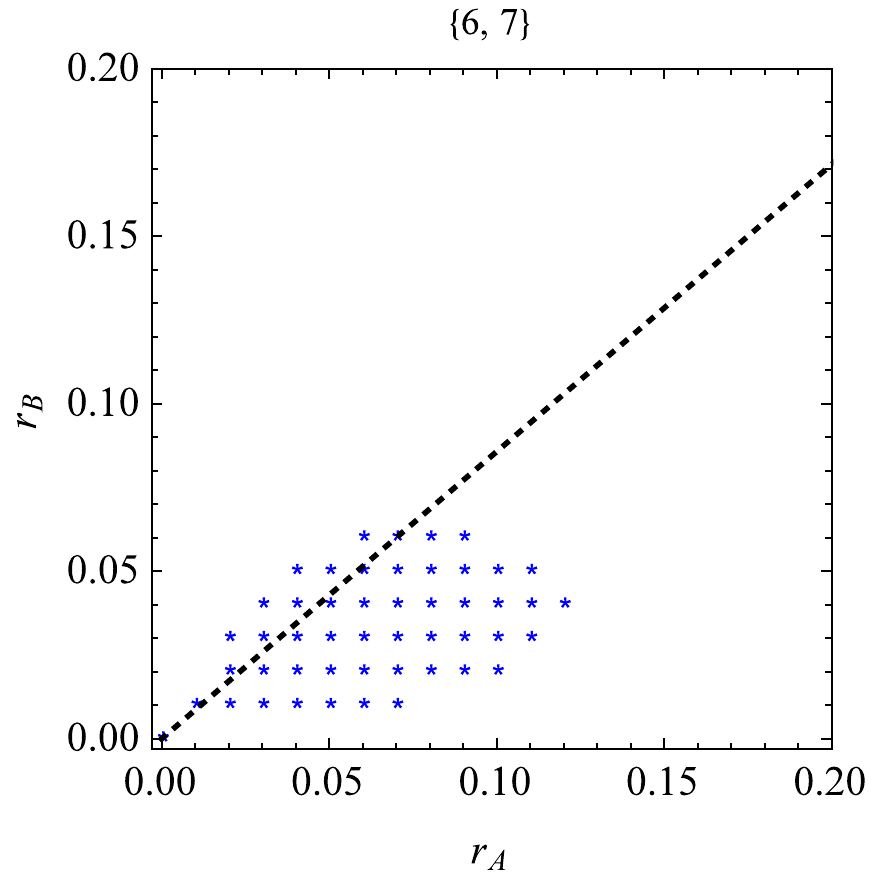}
\includegraphics[width=4cm]{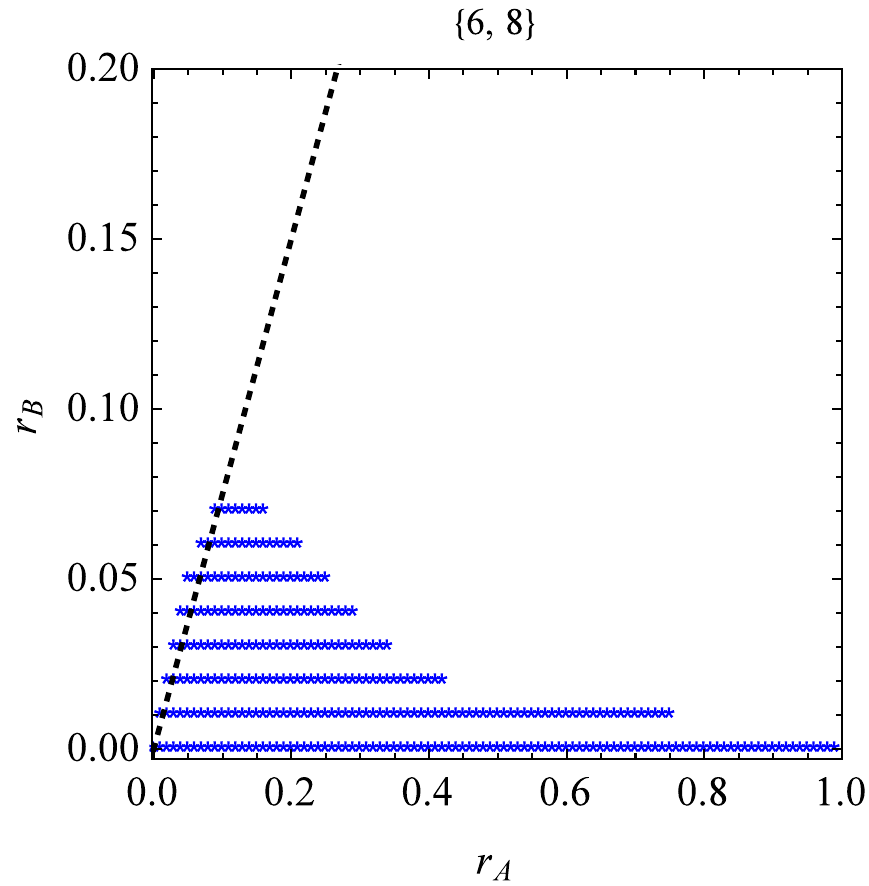}
\includegraphics[width=4cm]{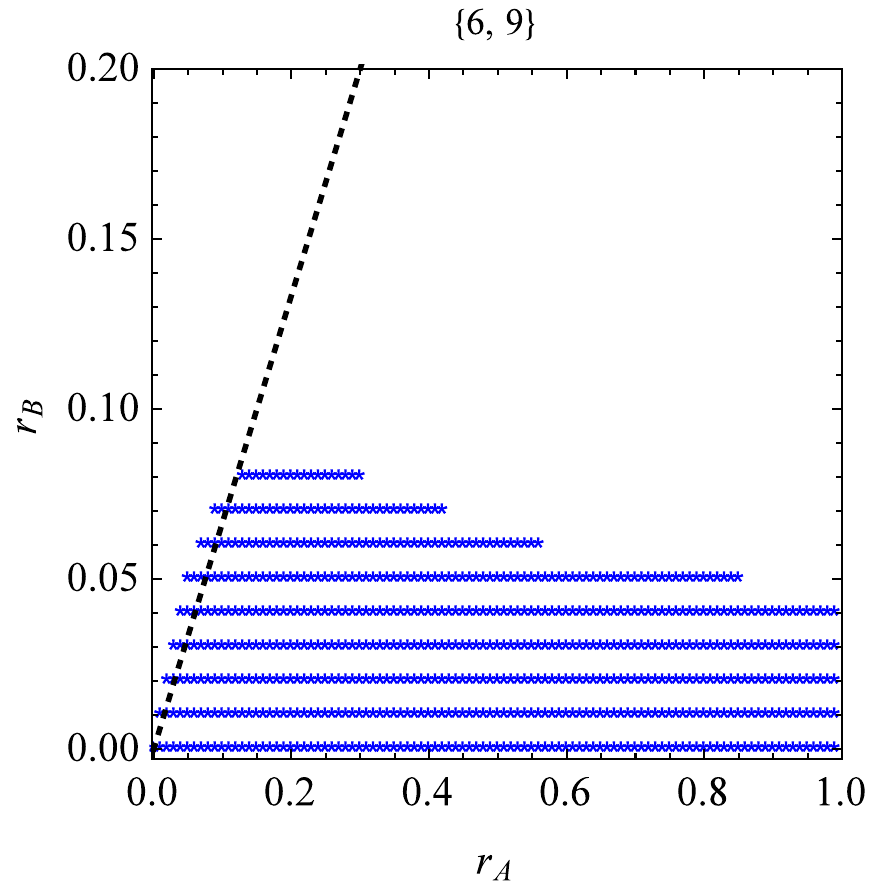}}
\caption{\label{fig:bAbBexample} The projection of the parameter scan on the $r_A$-$r_B$ plane in Case A for different $\{N_A,N_B\}$. The step size of the parameter scan is $\delta r_i=0.01 $.}
\end{figure}

\linespread{1}\begin{figure}[!h]
  \centering%
    \includegraphics[width=8cm]{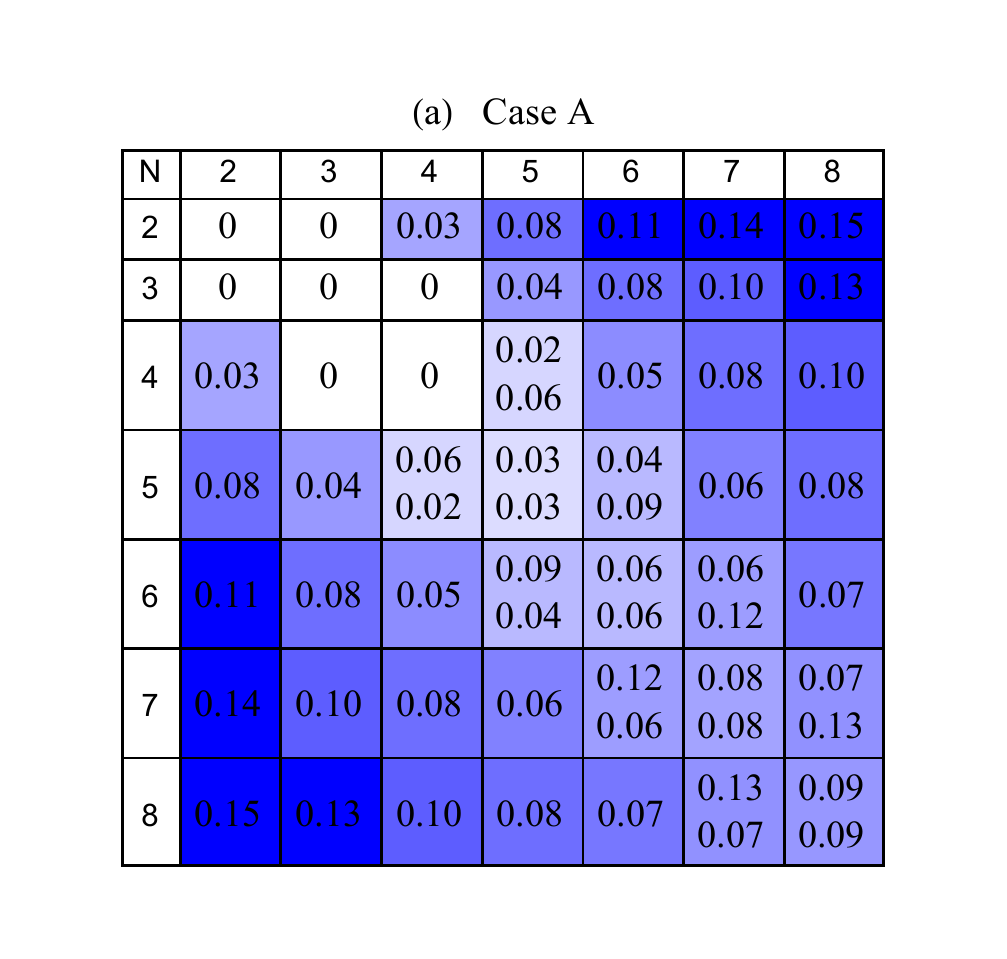}
    \includegraphics[width=8cm]{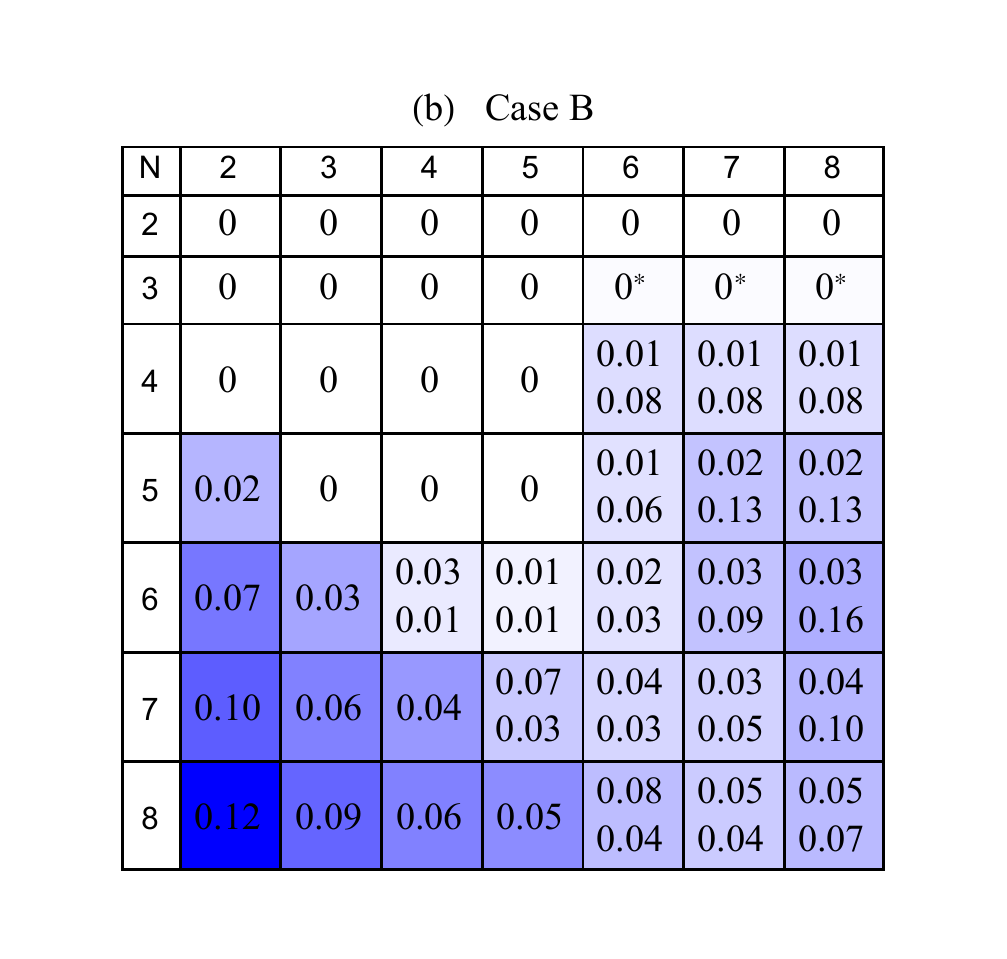}\\[-7mm]
    \includegraphics[width=8cm]{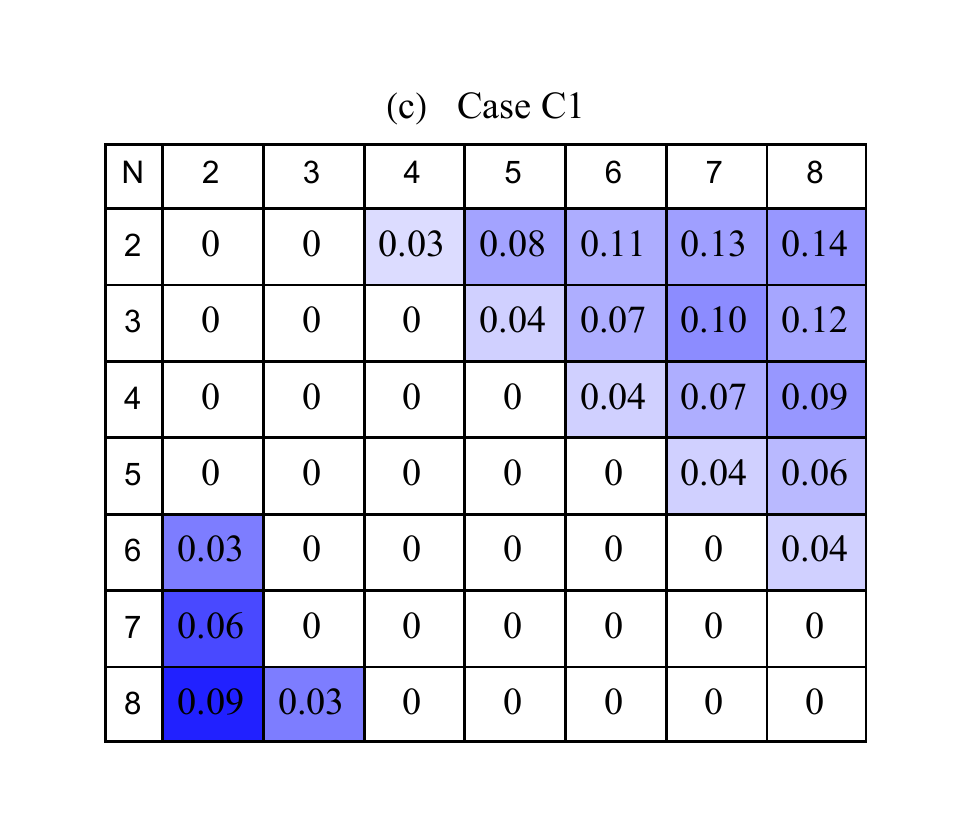}
    \includegraphics[width=4.5cm]{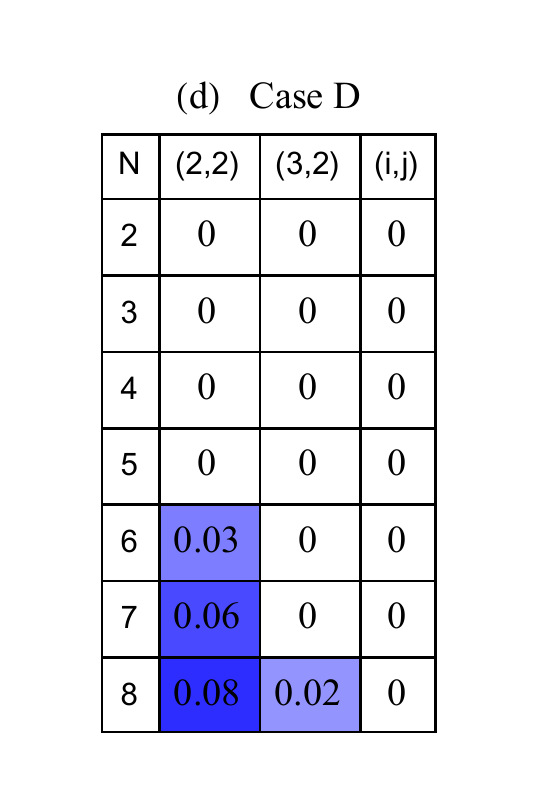}
\caption{\label{fig:PTbi} The upper bounds of $r_i$ where 0 means no solutions. For the first three cases they are functions of $N_A$ (row) and $N_B$ (column);  the last one is a function of $N_A$ (row) and $(N_B, N_C)$ (column). A single number gives the upper bound on the $r_i$ with the largest $N_i$. Two numbers provide limits on $r_B$ (upper) and $r_A$ (lower). $0^*$ denotes marginal cases where the existence of solutions goes beyond our parameter scan accuracy.}
\end{figure}

We find that the distribution of solutions as a function of the $r_i$'s share similar features for all our benchmarks. For each $N_i$ set we scan over $r_i$ space with the step $\delta r_i=0.01$ for $0<r_i\leq1$. This step is comparable to the minimum matter contribution for $N_i\lesssim 10$. The projections on the $r_A$-$r_B$ plane for Case A with $N_A=6$ and $2\leq N_B\leq9$ are presented in Fig.~\ref{fig:bAbBexample}. In each panel the black dot line denotes $b_A=b_B$. This figure highlights the fact that it is a large hierarchy between $N_A$ and $N_B$ that helps most to achieve a SAFE. And when there is a hierarchy it is the $r_i$ of the larger gauge group that is bounded from above.

\linespread{1}\begin{figure}[!h]
  \centering%
{ \includegraphics[width=7cm]{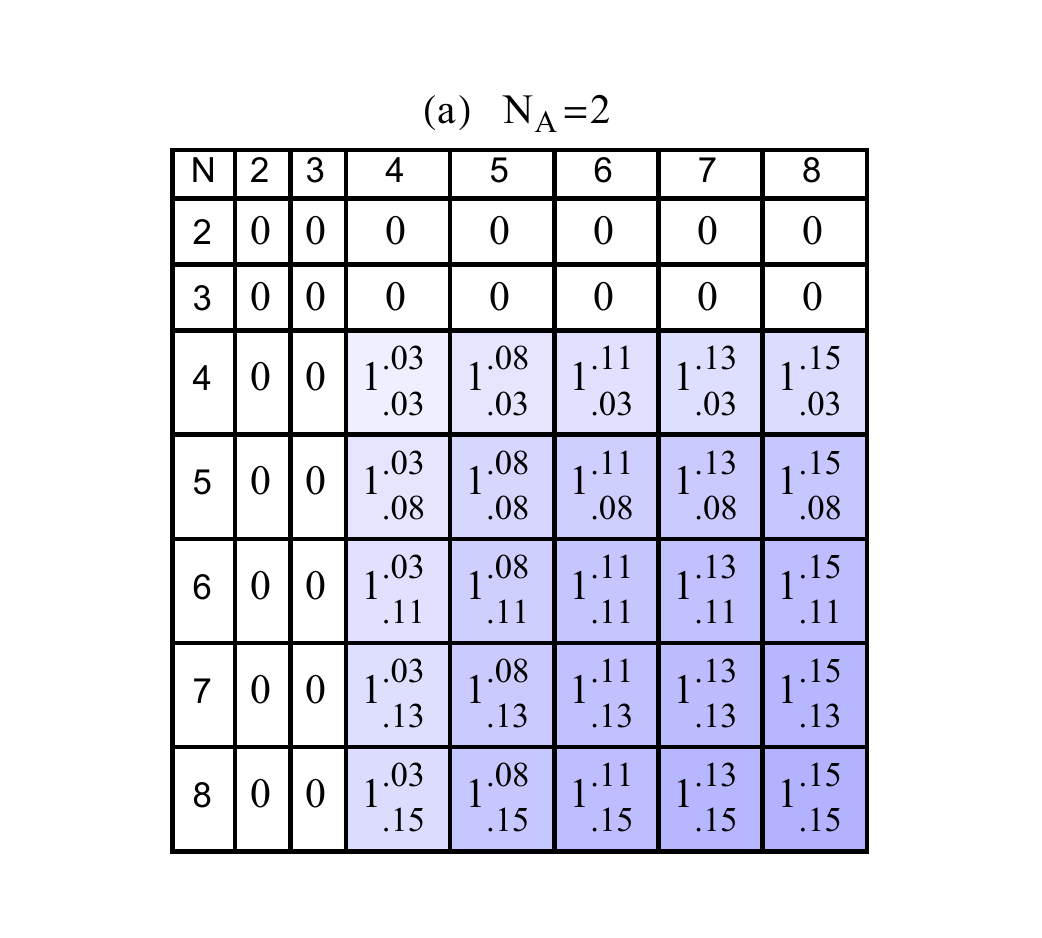}}
{ \includegraphics[width=7cm]{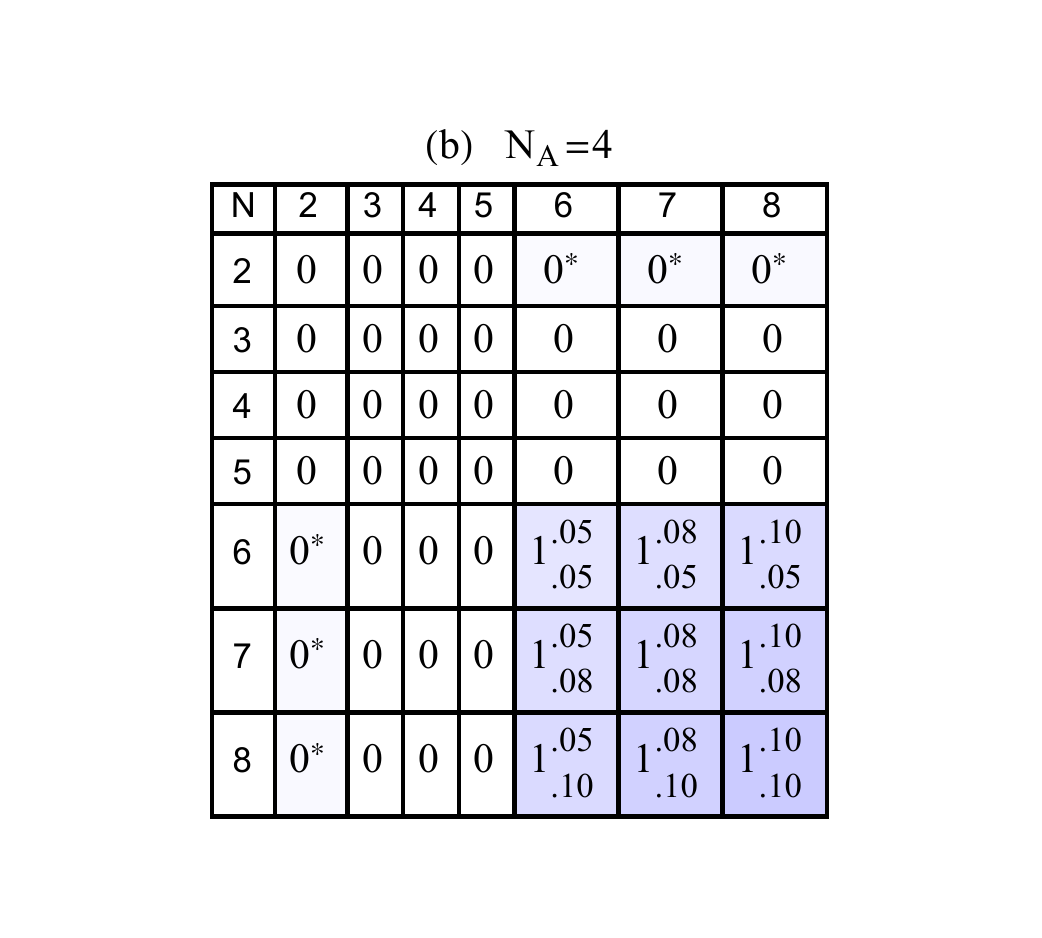}}\\[-2mm]
{ \includegraphics[width=7.2cm]{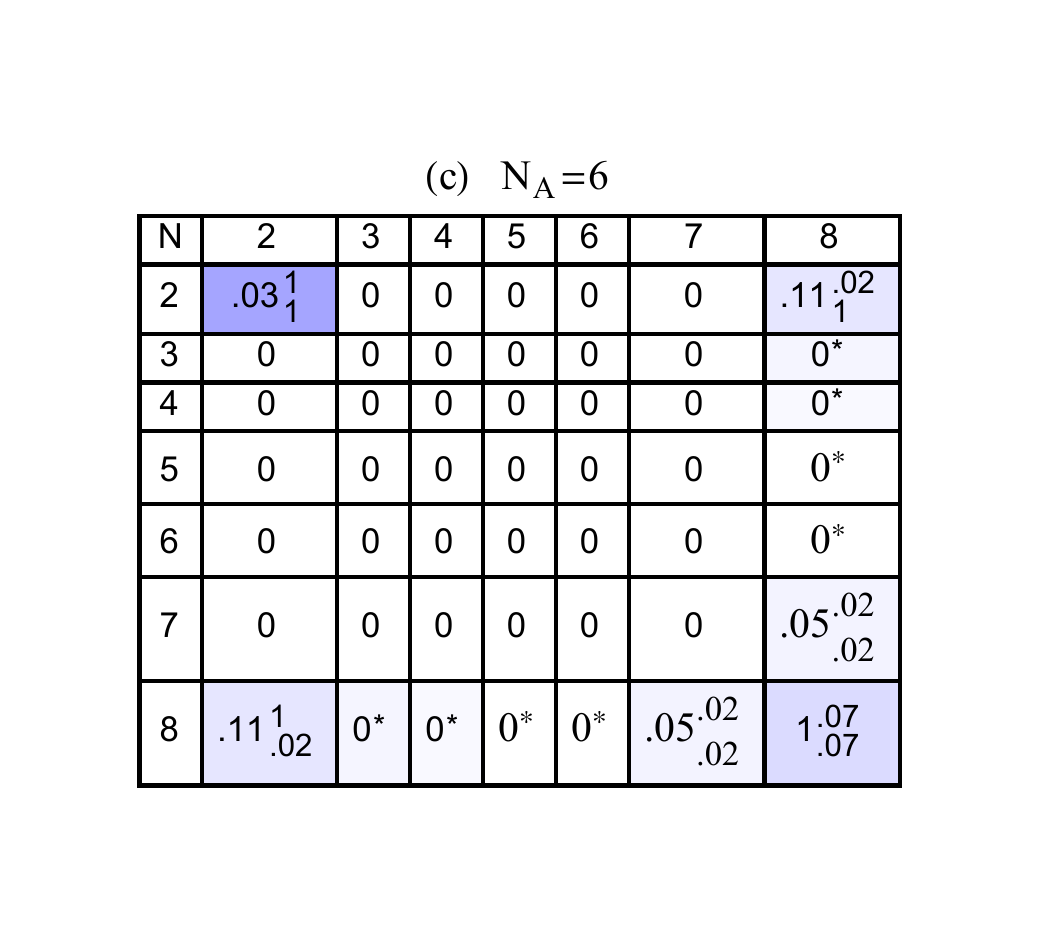}}
{ \includegraphics[width=6.8cm]{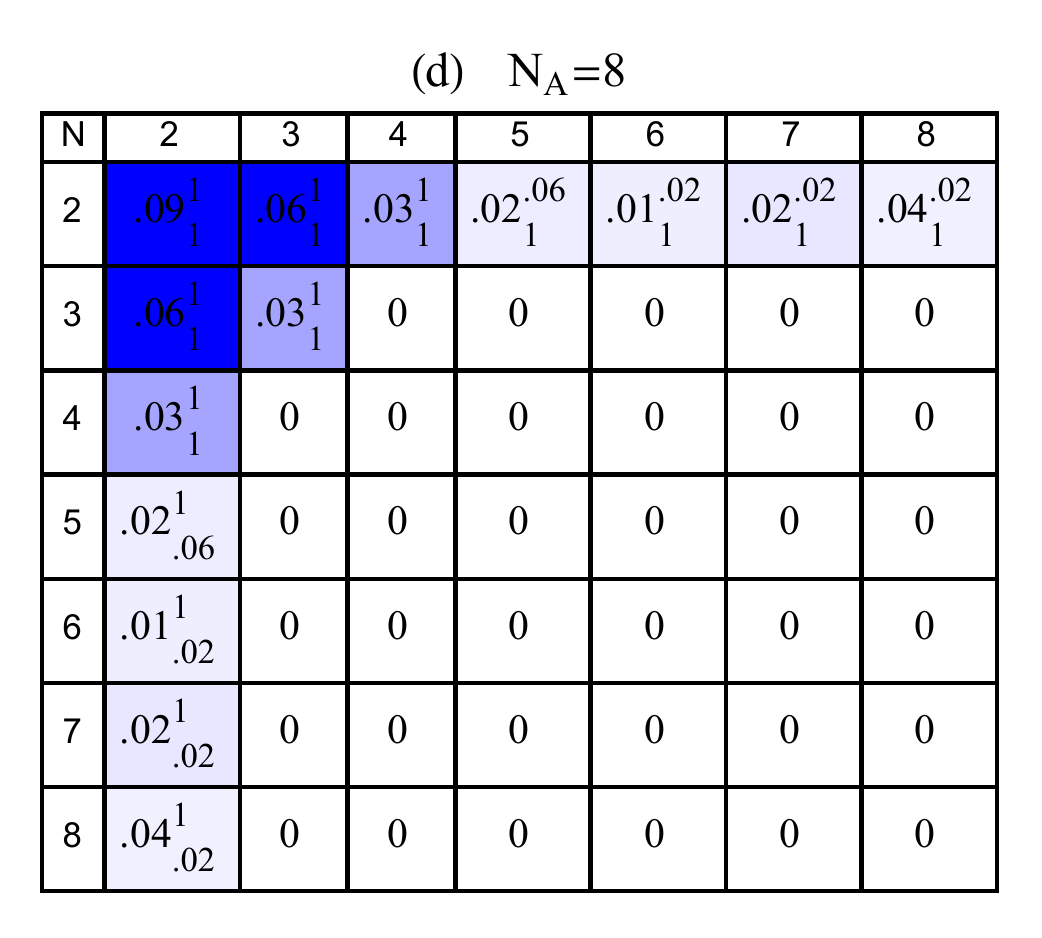}}
\caption{\label{fig:PTbiCaseC2} In Case C2 the upper bounds on $r_i$ for various $N_A$ as functions of $N_B$ (row) and $N_C$ (column). The three constraints are  presented with the notation $(r_A)^{r_C}_{r_B}$.}
\end{figure}

We present the upper bounds on $r_i$ for all our benchmark models in Fig.~\ref{fig:PTbi} and Fig.~\ref{fig:PTbiCaseC2}. This information can be used to constrain the matter content to achieve SAFEs. To illustrate the number fraction of viable points for each $N_i$ set, we use dark (light) blue for more (less) viable points. Fig.~\ref{fig:PTbi}(a) for Case A is symmetric under $N_A\leftrightarrow N_B$ and the general features mentioned above are quite clear. Well off the diagonal only the $r_i$ of the large gauge group is constrained and this constraint becomes more relaxed for increasing hierarchy between $N_A$ and $N_B$. For the near-diagonal elements there are upper bounds on both $r_B$ (upper) and $r_A$ (lower) for the two gauge $\beta$-functions. The $N_A=N_B=2$ case has a larger set of quartic couplings and we have checked that it does not yield a SAFE.

 We may briefly consider the fate of the fast running solutions, as we did for the simple gauge group. The vanishing of the linear terms in (\ref{eq:betaNANBr}) defines a boundary on the $r_A-r_B$ plane as follows,
\begin{eqnarray}\label{eq:dualB}
2+\frac{6}{r_Ab_{A,M}}\left(N_A-\frac{1}{N_A}\right)+\frac{6}{r_Bb_{B,M}}\left(N_B-\frac{1}{N_B}\right)=0\,.
\end{eqnarray}
The region below (above) the boundary features slow (fast) running, and the UVFP solutions in the two regions are related by a rescaling of the $r_i$ and $\lambda_i\to-\lambda_i$. In this Case A we find that the boundary (\ref{eq:dualB}) and thus all fast running UVFP solutions are outside of the physical region $0<r_i\leq1$.

For Case B in Fig.~\ref{fig:PTbi}(b), since the two scalars are charged differently under $SU(N_A)\times SU(N_B)$, the pattern becomes asymmetric. The rows and columns denote the common gauge group $SU(N_A)$ and $SU(N_B)$ respectively. When $N_A\geq N_B$ we see the similar pattern as Case A in the lower left part of the table but with a smaller viable parameter space. In the upper right corner, i.e. $N_B>N_A$, the common group is small and then for the $(N_A,1)$ scalar it is difficult to obtain solutions.

In Fig.~\ref{fig:PTbi}(c) we present the bounds for Case C1 with $Z_2$ symmetry, with row and column for $SU(N_A)$ and $SU(N_B)$ respectively. The $N_A>N_B$ region, the lower left corner, now has a more stringent constraint on $r_A$ compared to Cases A and B. This is due to enhanced pure quartic terms in the $\beta$-functions of (\ref{eq:beta31}). For the $N_B>N_A$ region, the upper right corner, there are more solutions compared to Case B since the two copies of $SU(N_B)$ enhance the gauge-quartic terms. Here the constraint on $r_B$ applies to both of the large $SU(N_B)$ gauge groups. For the special case $N_A=4, N_B=N_C=2$ where we see zero solutions, one more coupling $\lambda_\epsilon$ in (\ref{eq:potential31LC}) gets involved. Given that its $\beta$-function is proportional to $\lambda_\epsilon$, its UVFP is at zero, and so whether or not it is stable it cannot alter the lack of a UVFP in the other couplings.

For Case D we only find a small number of $(N_A, N_B, N_C)$ values with viable solutions, as shown in Fig.~\ref{fig:PTbi}(d). Here we assume $N_A$ (row) is the largest while $(N_B, N_C)$ (column) has $N_B\geq N_C$. The paucity of solutions here is basically due to the appearance of a $4N_AN_BN_C\bar\lambda_d^2$ term in $\beta_{\lambda_d}$. Again the extra coupling $\lambda_\epsilon$ in (\ref{eq:CaseDLC}) for the special case $N_A=4, N_B=N_C=2$ does not affect the lack of a UVFP.

Finally we turn to Case C2. It depends on all three $N_A, N_B, N_C$ and the results cannot be summarized in one 2D plane. But we do find that the constraints when $N_B=N_C$ are quite similar to Case C1 in Fig.~\ref{fig:PTbi}(c). The general upper bounds on $r_A, r_B, r_C$ for various $N_A$ are displayed in Fig.~\ref{fig:PTbiCaseC2} as functions of $N_B$ (row) and $N_C$ (column). We present these limits using the notation $(r_A)^{r_C}_{r_B}$. From the four tables one can see that solutions tend to appear when some hierarchy develops between the three values $N_A,N_B,N_C$. Among the possibilities, a hierarchy with a large common gauge group is the most efficient. And it can be seen that the upper bound on $r_i$ is typically relaxed or nonexistent ($=1$) in those cases where the associated $N_i$ is small relative to some other $N_j$.

\subsection{$\bar\lambda_j$ values from the parameter scan}

Next we show results for the values of the quartic couplings at the UVFPs. We define $\bar\lambda_j\equiv\lambda_j/g^2_i$ where $g_i$ is the coupling of the largest gauge group. We saw in previous section that this coupling runs most slowly in the UV (has the smallest $b_i$) and thus is the largest gauge coupling.

\linespread{1}\begin{figure}[!h]
  \centering%
{ \includegraphics[width=5.3cm]{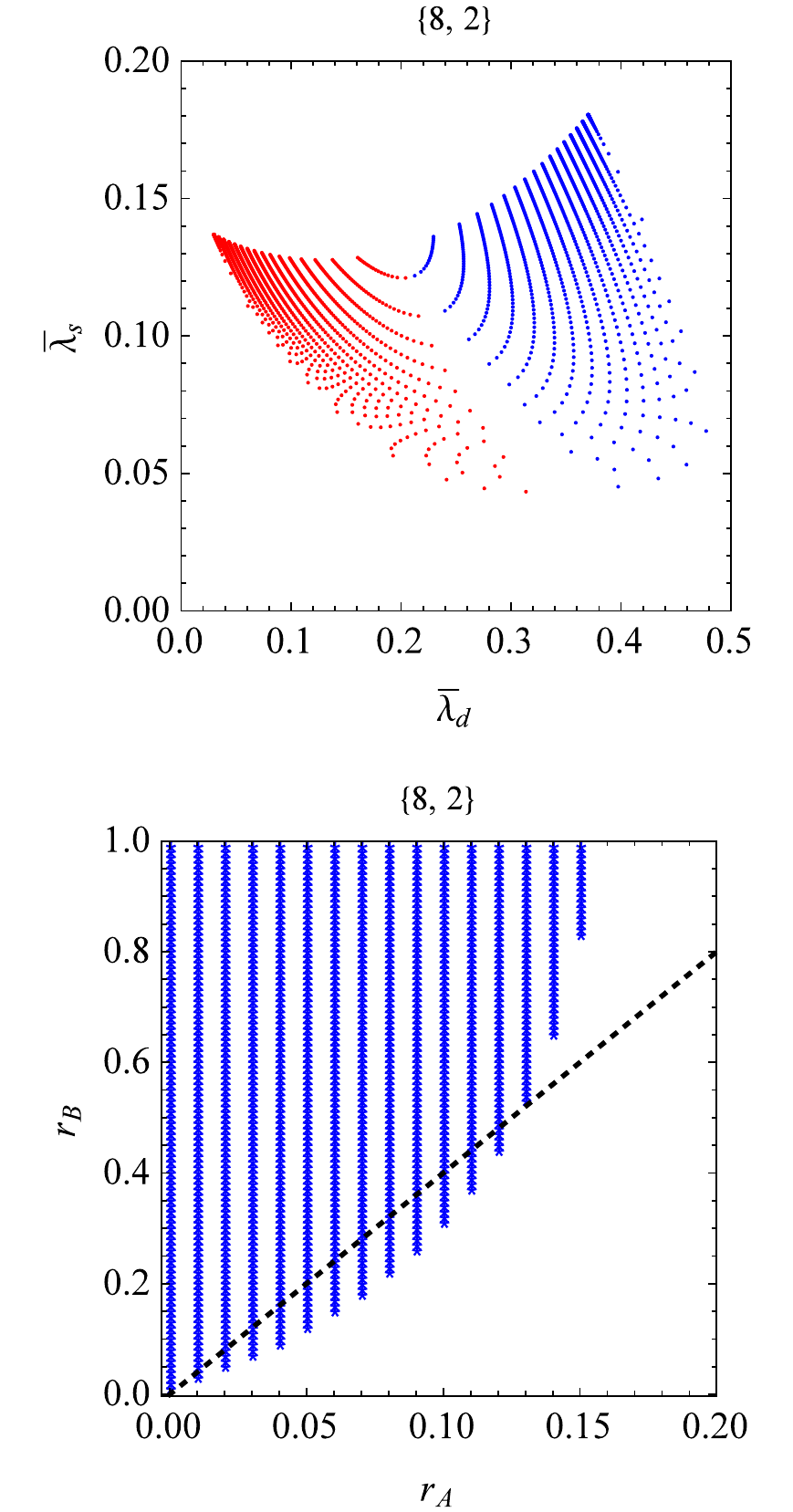}
\includegraphics[width=5.3cm]{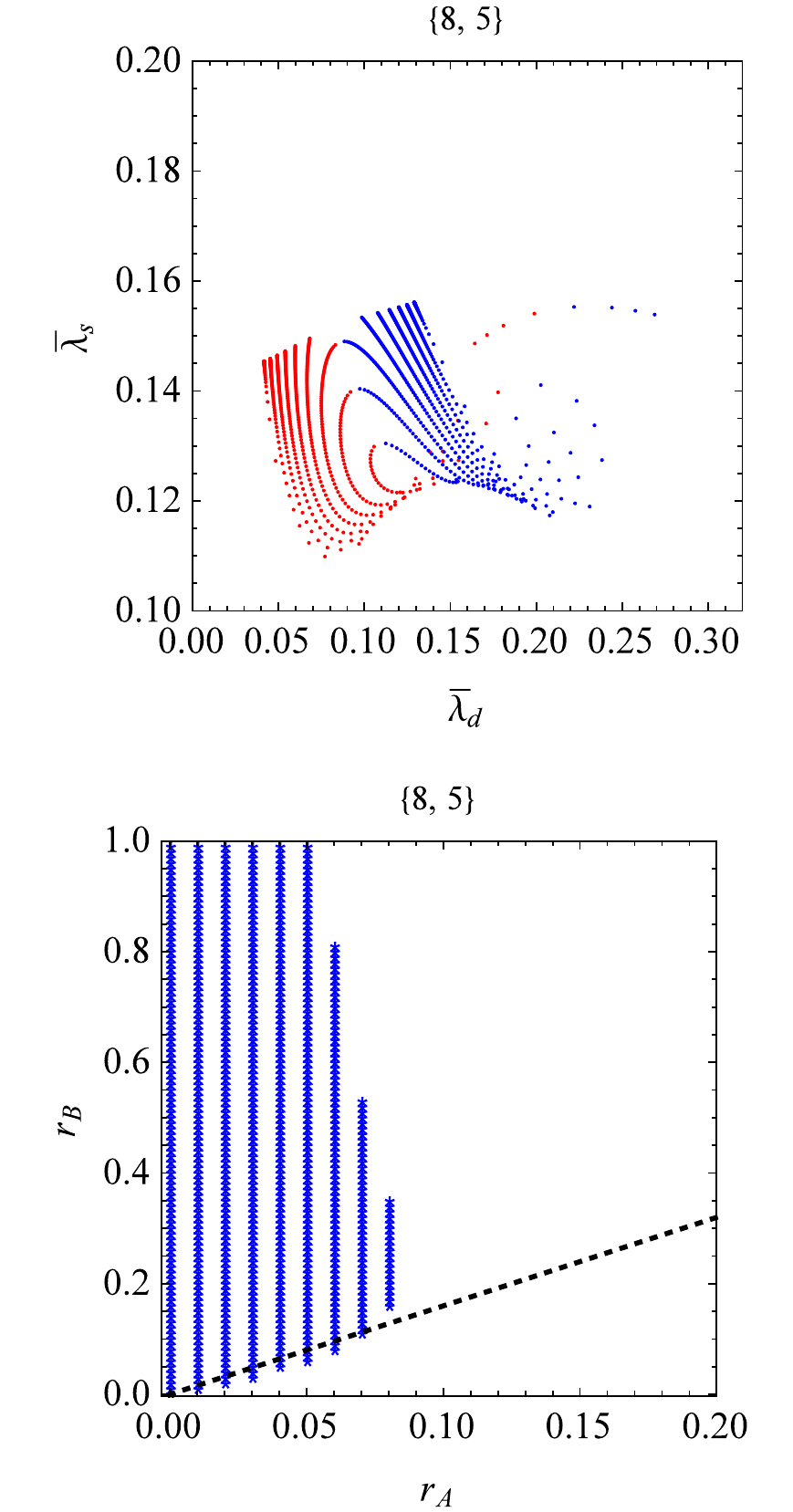}
\includegraphics[width=5.3cm]{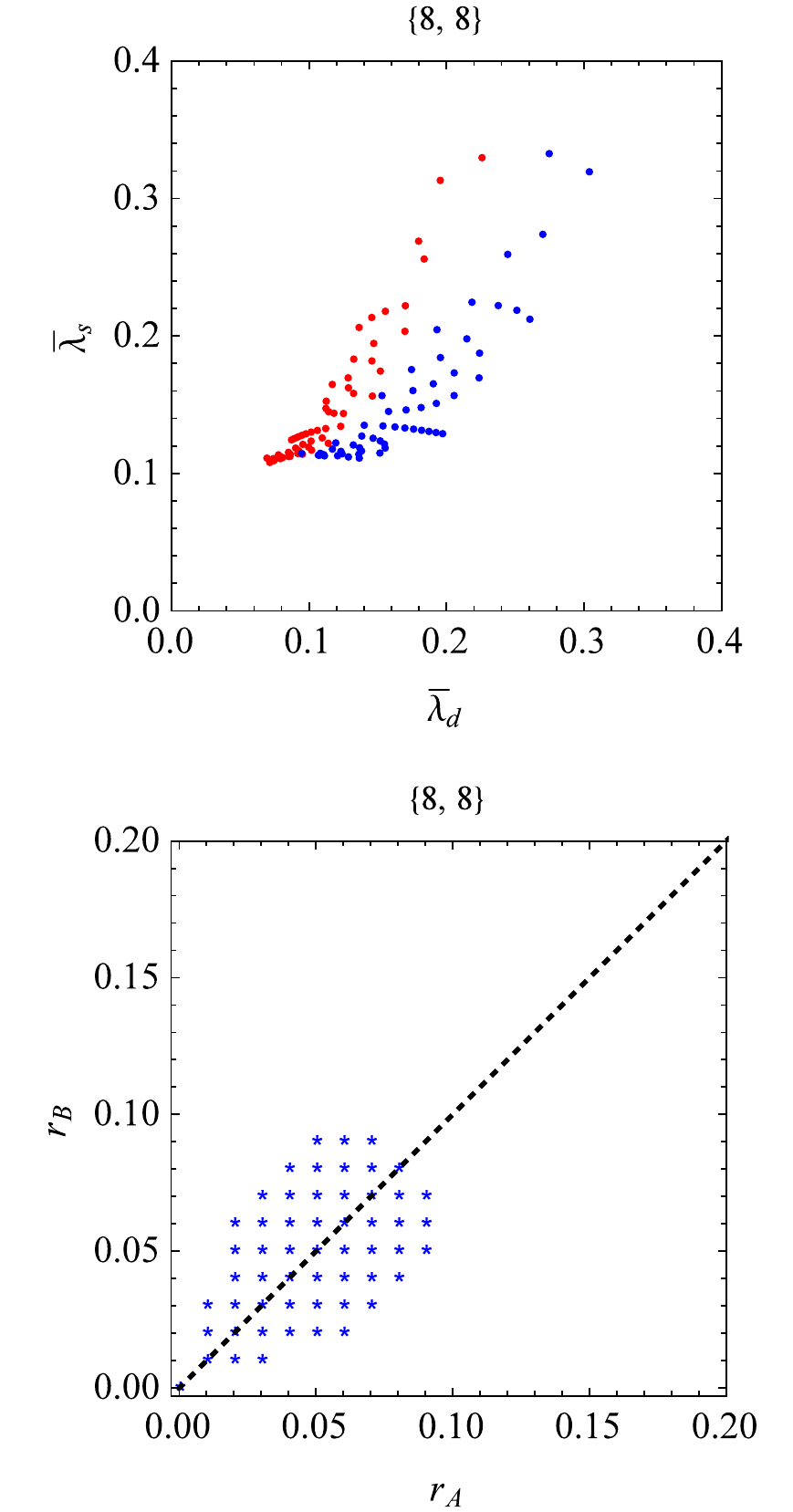}}
\caption{\label{fig:ladsCaseA} The projection of the parameter scan on the $\bar\lambda_d-\bar\lambda_s$ (first row) and $r_A$-$r_B$ (second  row) planes for different $\{N_A, N_B\}$. The quartic couplings are normalized by the largest gauge coupling. The red and blue dots represent stable and unstable UVFPs respectively. Note that some characteristics of these plots are determined by the step size of the parameter scan.}
\end{figure}

We start from the simplest Case A with only two quartic couplings. In Fig.~\ref{fig:ladsCaseA}, for some typical $(N_A, N_B)$, the first row shows the projection of the parameter scan on the $\bar\lambda_d$-$\bar\lambda_s$ plane, while the second row shows the $r_A$-$r_B$ projection for comparison. Among all UVFP of (\ref{eq:betaNANBr}) we depict the stable and unstable solutions by red and blue dots respectively. The situation is clearest for the left plots where the ratio $N_A/N_B$ is the greatest. For each $(r_A, r_B)$, there are always a pair of solutions, one stable and one unstable with smaller and larger $\bar\lambda_d$ respectively. With decreasing $r_A$ we go through different arcs from inside out, where the arc length depends on the number of viable $r_B$. In $r_A\to0$ limit, the solutions become independent of $r_B$ and reach the corners of the red and blue regions that possess the largest distance between stable and unstable UVFPs. When $N_A, N_B$ are similar both gauge couplings play significant roles and the solution pattern becomes more involved.

 The unstable solution in each case is actually a saddle point, with one direction UV attractive and the other one repulsive. Also, at least for $2\leq N_i\leq 8$, we find that the quartic couplings at the UVFPs are positive and typically of order 0.1 or 0.2 times the largest gauge coupling. The stability of tree level potential demands the conditions
\begin{eqnarray}\label{eq:CaseAVS}
\bar\lambda_d+\bar\lambda_s>0,\quad
2\bar\lambda_d+\bar\lambda_s>0\,,
\end{eqnarray}
but here they put no further constraint.

In comparison to these slow running UVFPs the unphysical fast running UVFPs again come in pairs, but one is a saddle point and the other is completely unstable. Another curiosity occurs when one of the $N_i$ is very large, e.g. $N_A\geq26$ and $N_B=2$. Then four slow running UVFPs can occur, one stable, two saddle points, and one completely unstable. The two new UVFPs correspond to a large $\bar\lambda_s>0$, with which the coefficient of linear $\bar\lambda_d$ term in $\beta_{\bar\lambda_d}$ becomes positive and the root of $\bar\lambda_d$ is negative. The fast running version of these UVFPs would be characterized by the same four types, which is more interesting here because one is stable. But at least for the cases we have considered the fast running solutions are outside the physical range of the $r_i$'s, and they produce tension for Yukawa couplings and vacuum stability.

\linespread{1}\begin{figure}[!h]
  \centering%
{ \includegraphics[width=5.22cm]{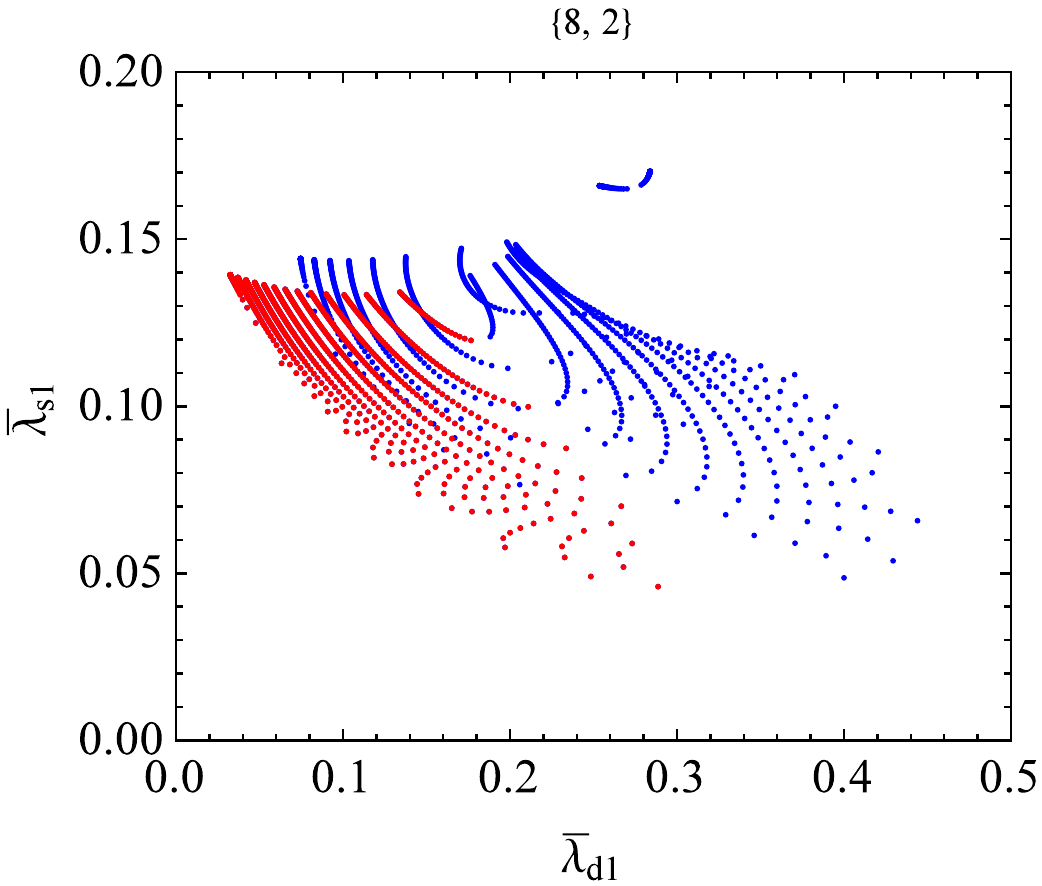}
\includegraphics[width=5.1cm]{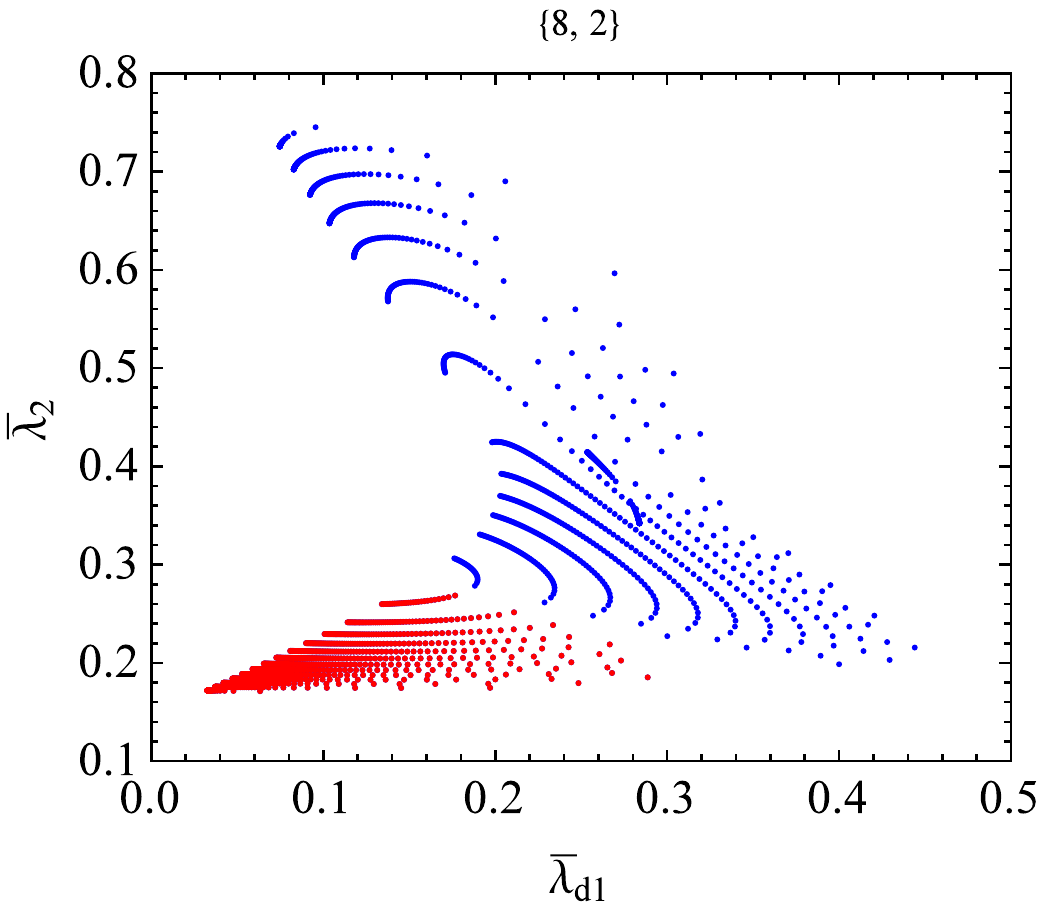}
\includegraphics[width=5.3cm]{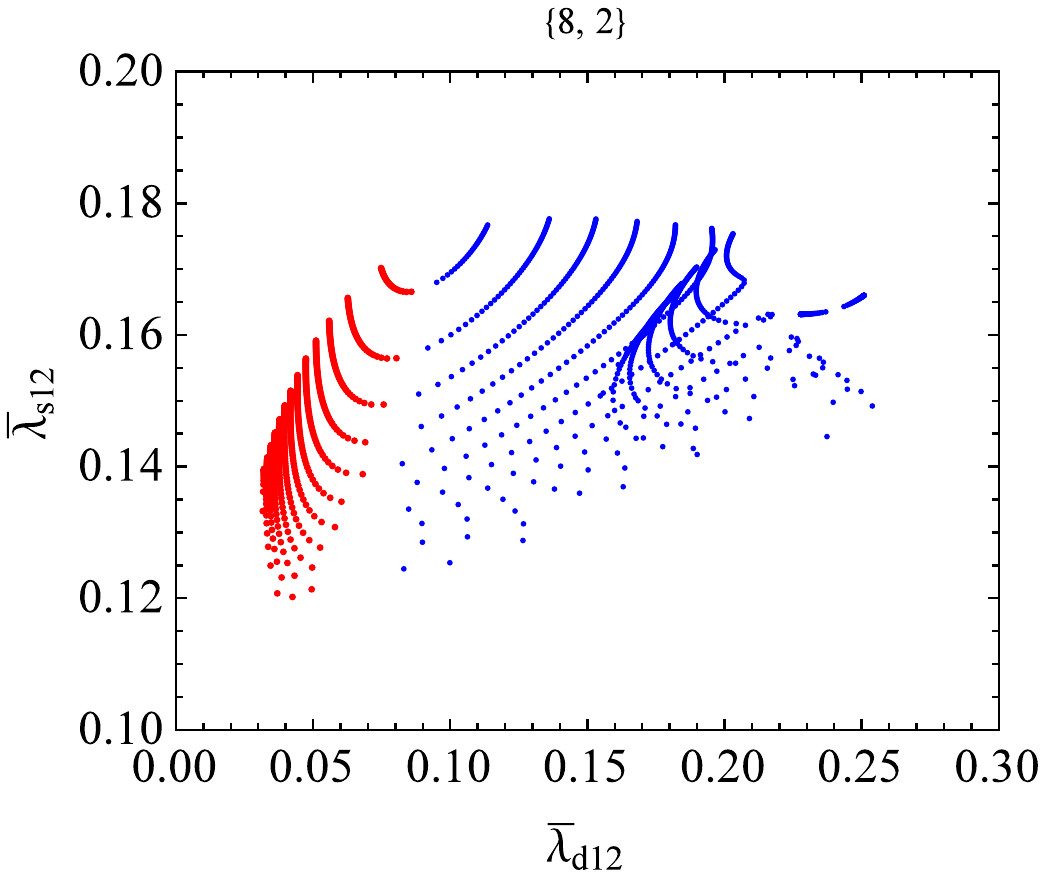}}
\caption{\label{fig:ladsCaseB} Projection of the parameter scan on some coupling planes for Case B with $N_A=8, N_B=2$.}
\end{figure}

For Case B with five quartic couplings we project the higher dimensional space onto three 2D planes. In Fig.~\ref{fig:ladsCaseB} we show the case $N_A=8, N_B=2$. Compared with the counterpart in Case A we see a similar pattern of stable and unstable UVFP pairings on the $\bar\lambda_{d1}$-$\bar\lambda_{s1}$ plane. For some $r_i$ there are four UVFPs and the additional pair of solutions are saddle points. They correspond to different $\bar\lambda_2$ as shown on $\bar\lambda_{d1}$-$\bar\lambda_2$ plane. The mixing couplings $\bar\lambda_{s12}$ and $\bar\lambda_{d12}$ are both positive and away from zero. They make considerable positive contribution to $\beta_{\lambda_{d1}}, \beta_{\lambda_{s1}}, \beta_{\lambda_{2}}$, causing the number of solutions to decrease.

\linespread{1}\begin{figure}[!h]
  \centering%
{ \includegraphics[width=5.6cm]{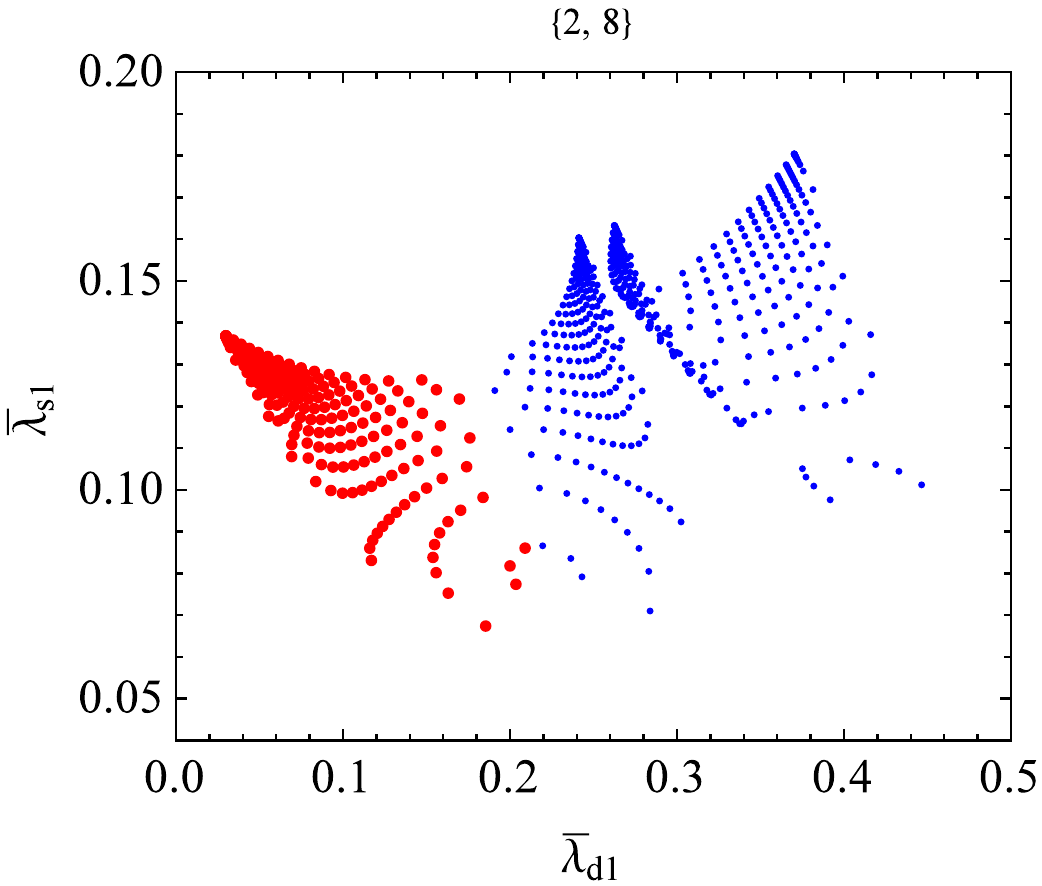}
\includegraphics[width=5.6cm]{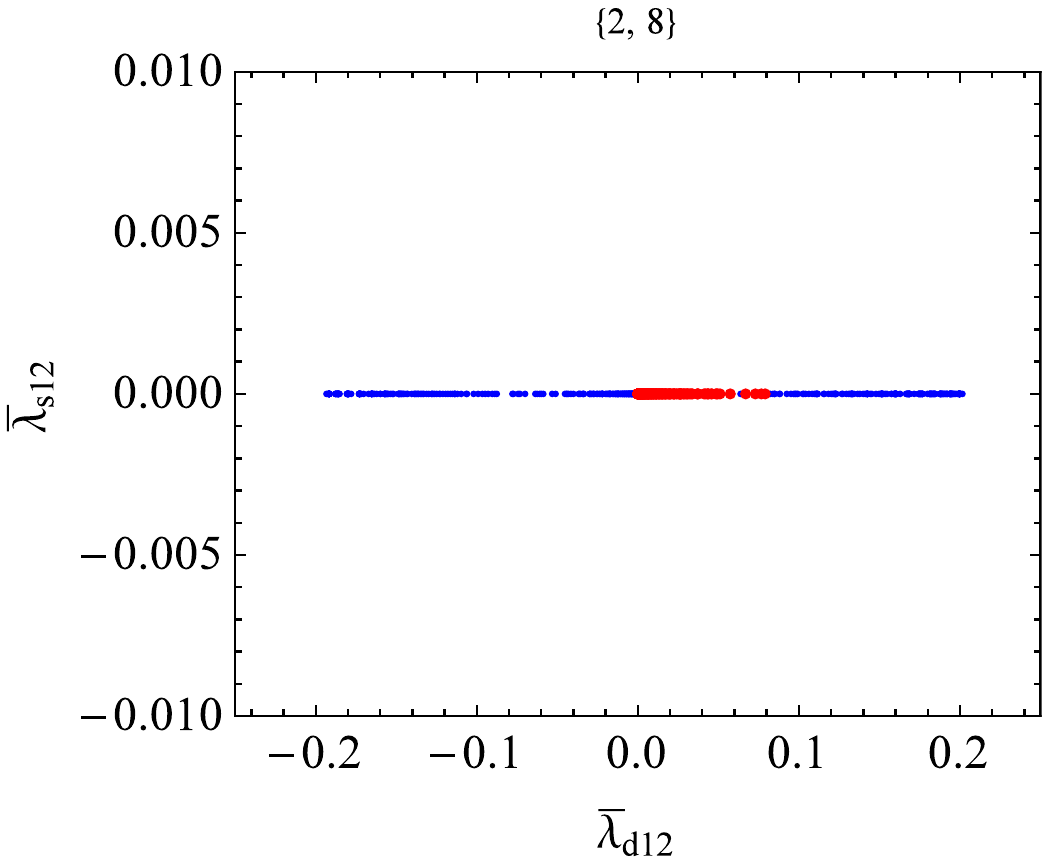}
\includegraphics[width=4.7cm]{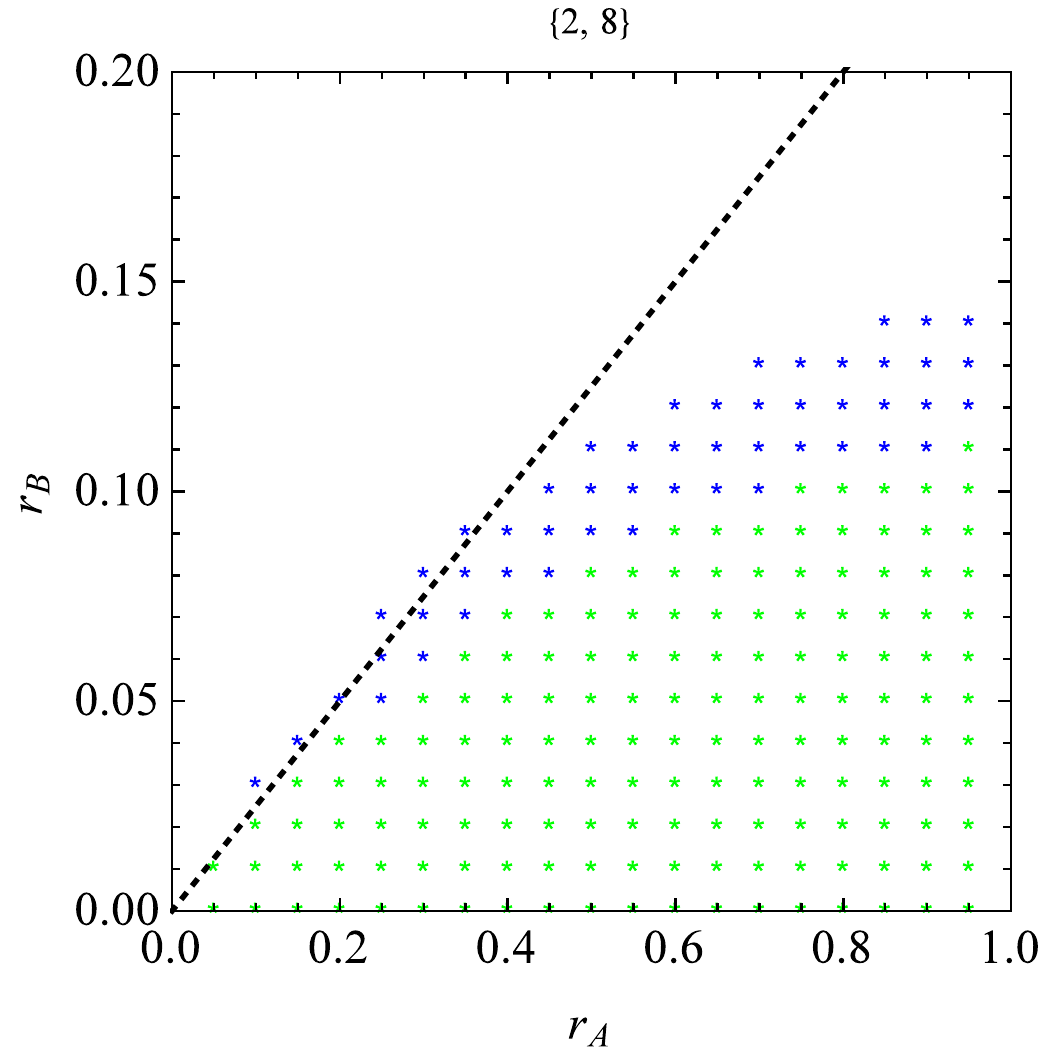}}
\caption{\label{fig:ladsCaseC} Projection of the parameter scan on some coupling planes for Case C1 with $N_A=2, N_B=8$. The right panel is a $r_i$ projection, where blue and green denote the points with 2 and 4 UVFPs respectively.}
\end{figure}

For Case C1 the large common group case $N_A>N_B$ has quite similar features to Case A. Given the dominance of the common gauge group there are two UVFPs for each viable $r_i$ and the one with the smaller $\bar\lambda_{d}$ is UV stable. In the small common gauge group case, $N_A<N_B$, some new types of solutions emerge. For illustration we present the UVFPs for $N_A=2, N_B=8$ in the left and middle panels of Fig.~\ref{fig:ladsCaseC}. For some $r_i$ there are again an extra pair of UVFPs at saddle points. They possess a large $\bar\lambda_{d1}$ (left) and a negative $\bar\lambda_{d12}$ (middle). The corresponding $r_i$ with four UVFPs are denoted by the green dots in the right panel.

With a large positive $\bar\lambda_{d1}$ we find that the mixing coupling $\bar\lambda_{d12}$ can be negative, but then the coefficient of $\bar\lambda_{d12}$ in $\beta_{\bar\lambda_{d12}}$ is positive and the solution becomes unstable. Mixing couplings are usually positive for stable UVFPs, but a new feature we see here is that they can be close to zero. This is due to the suppressed pure gauge terms in the $\beta$-functions of the mixing couplings, which only receives a small contribution from the common gauge group (it is 0 for $N_A=2$ case). Finally the general picture of UVFPs for Case C2 without the $Z_2$ symmetry is similar to Case C1.

\linespread{1}\begin{figure}[!h]
  \centering%
{ \includegraphics[width=7cm]{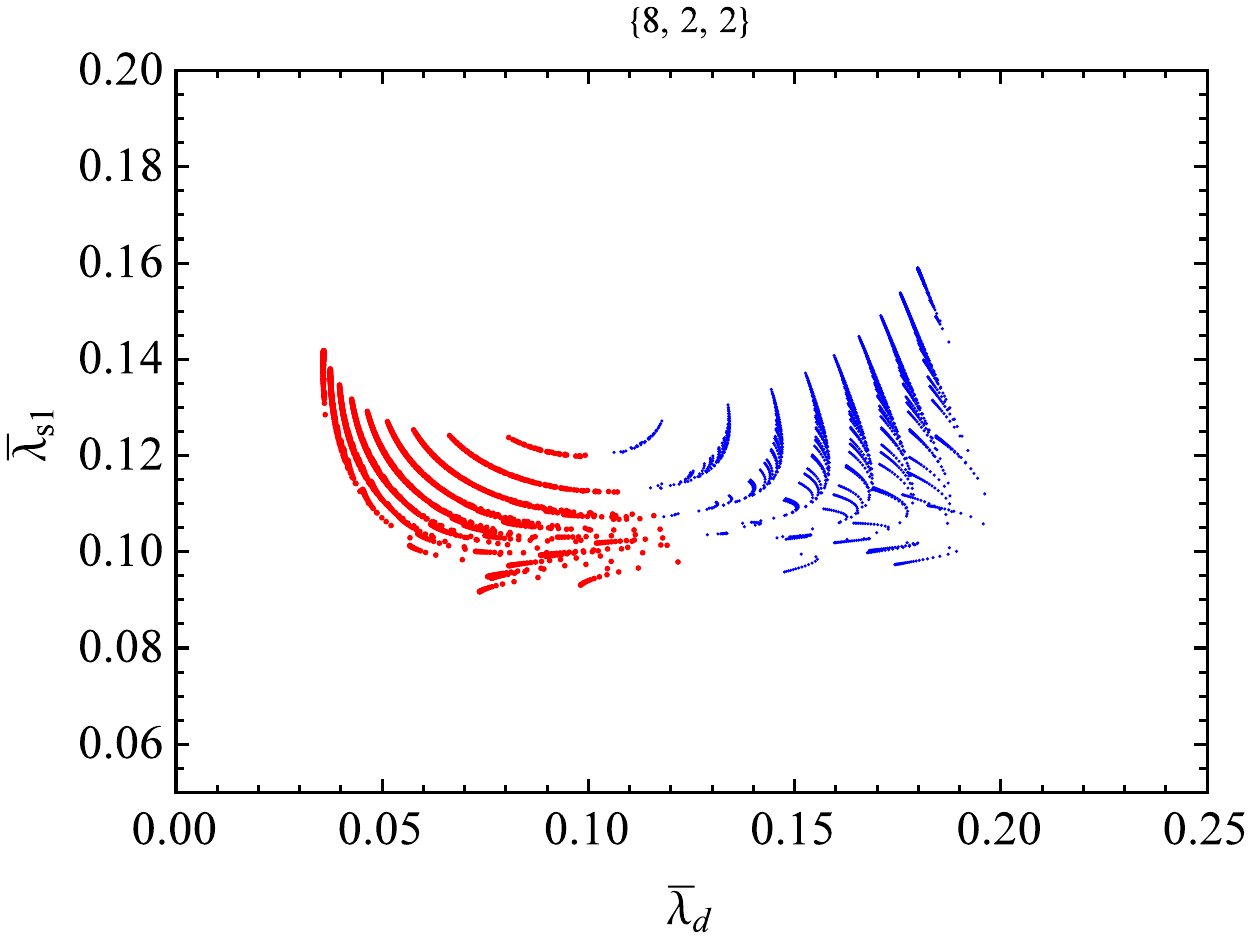}
\includegraphics[width=7cm]{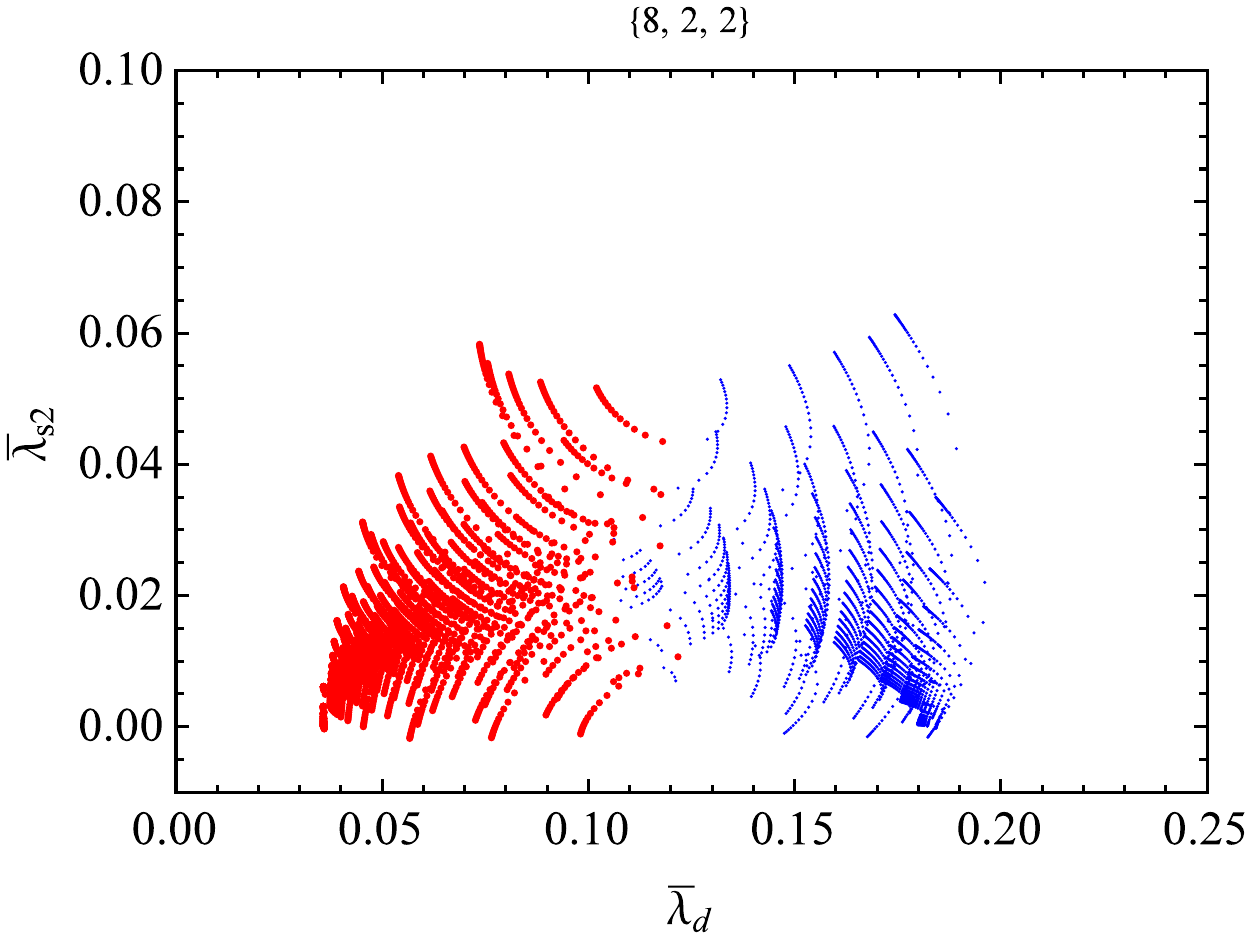}}
\caption{\label{fig:ladsCaseD} Projection of the parameter scan on some coupling planes for Case D with $N_A=8, N_B=2, N_C=2$.}
\end{figure}

Case D has four quartic couplings, one double trace and three that are single trace. We depict the projections $\bar\lambda_{s1}$-$\bar\lambda_d$ and $\bar\lambda_{s2}$-$\bar\lambda_d$ in Fig.~\ref{fig:ladsCaseD} for $N_A=8$, $N_B=N_C=2$. The typical feature is reflected on the range of single trace couplings at UVFPs. We find that the coupling with a single trace associated with the largest gauge group $\bar\lambda_{s1}$ has comparable size with other couplings at UVFPs, while those associated with small gauge groups, $\bar\lambda_{s2}$ or $\bar\lambda_{s3}$, could be close to zero or even slightly negative. Again this is determined by the dominant pure gauge terms in the $\beta$-functions.

\section{The Simplest Model}
\label{models}

As a general feature of the previous results, when a hierarchy in the sizes of the different gauge groups helps to achieve SAFEs, the gauge coupling associated with the largest group is constrained to run quite slowly. A small ratio $r_i=b_i/b_{i,M}$ requires a sufficient number of fermions. We first check whether some number of chiral fermions gauged under $SU(N_A)\times SU(N_B)\times SU(N_B)$ could work. We assume the fermion content
\begin{align}\label{eq:CaseCFermion}
\Psi_L:\,(N_A,N_B,1),\quad
\Psi_R:\,(N_A,1,N_B),\quad Q_L:\,(1,\bar{N}_B,N_B),
\end{align}
with $n_F$ copies of $\Psi_L+\Psi_R$ and $n_Q$ copies of chiral fermions $Q_L$. To be anomaly free when $N_B>2$ we need an integer ratio $n_F/n_Q=N_B/N_A$. (For $N_B=2$ we only need $n_FN_A+n_QN_B$ to be even \cite{Witten:1982fp}.) The $\beta$-function coefficients of two gauge couplings are
\begin{align}
b_A=-\frac{11}{3}N_A+n_F\frac{2N_B}{3}+b_{A,s},\quad
b_B=-\frac{11}{3}N_B+n_F\frac{2N_A}{3}+b_{B,s},
\end{align}
if we use $n_Q=N_A n_F/N_B$. $b_{i,s}$ is the scalar contribution and for instance $b_{A,s}=N_B/3, b_{B,s}=N_B/6$ for Case C. Since the scalar contributions are small we need $n_F$ sufficiently large to render $b_i$ of the largest gauge group small for SAFEs as in Fig.~\ref{fig:PTbi}. On the other hand, $n_F$ is bounded from above by the requirement that all gauge couplings are asymptotically free, i.e. $b_A, b_B<0$. It turns out that no $n_F$ may satisfy both requirements. The alternative then is to introduce the appropriate number of fermions that only transform under the large gauge group.

Two low scale unification models with a long history in the literature are both based on a product of three gauge groups. One is the triunification model based on $SU(3)\times SU(3)\times SU(3)$ \cite{triu} and the other is the Pati-Salam model $SU(4)\times SU(2)_L\times SU(2)_R$ \cite{Pati:1974yy}. Our results show that the former cannot be SAFE and so we turn to the latter. In this case of all the SAFEs that we have found there is only one that is of relevance. From the results for Case A we find that we can add a single scalar $\Phi$ transforming as $(4,2,1)$. We choose $(4,2,1)$ rather than $(4,1,2)$ to ensure that $\Phi$ will yield the SM Higgs doublet.

As we have just discussed, the constraint on the $SU(4)$ $\beta$-function from Fig.~\ref{fig:PTbi}, $|b_4|\lesssim0.44$, requires additional fermions. Thus in addition to the $n_F$ families of standard fermions $F_{L/R}$ we have a number $n_f$ of Dirac fermions $f_{L/R}$ transforming only under $SU(4)$. These fermions are vector-like, they can have mass without breaking the gauge symmetries. These masses are additional parameters in the model. The particle content is then as shown in Table~\ref{tab:matterIR}. Upon the breakdown $SU(4)\to SU(3)$ we see that the model predicts a colored scalar doublet in addition to the Higgs doublet.
\begin{table}[h]
\begin{center}
\caption{Matter fields in the simplest model.}
\label{tab:matterIR}
\begin{tabular}{c||c|c|c|c}
\hline\hline
Fields & Number & $SU(4)$ & $SU(2)_L$ & $SU(2)_R$ \\
\hline
$F_L$  & $n_F$ & 4 & 2 & 1\\
$F_R$  & $n_F$ & 4 & 1 & 2\\
$f_{L,R}$  & $n_f$ & 4 & 1 & 1\\
\hline
$\Phi$  & 1 & 4 & 2 & 1\\
\hline\hline
\end{tabular}
\end{center}
\end{table}

\linespread{1}\begin{figure}[!h]
  \centering%
    \includegraphics[width=15cm]{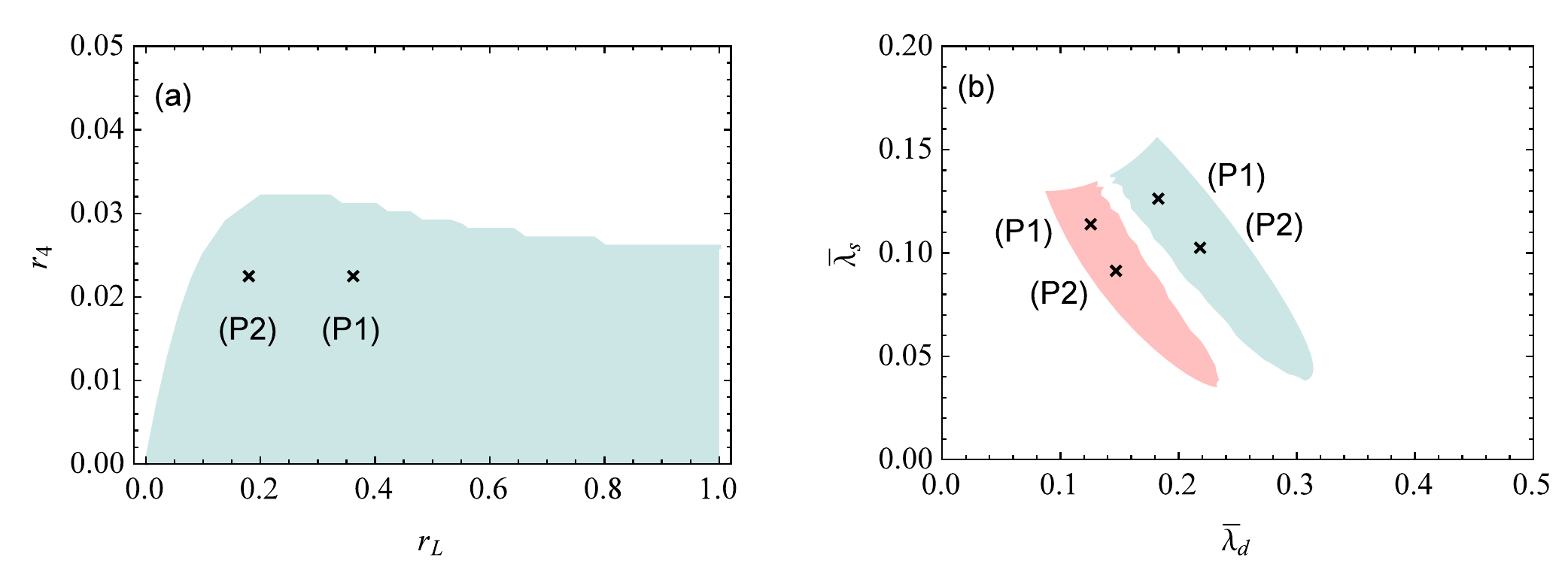}
\caption{\label{fig:NA4NB2CaseAplots} The two viable points in (\ref{pp12}) showing (a) the $\beta$-function coefficients and (b) the coupling ratios at the UVFPs.}
\end{figure}

The one loop $\beta$-functions are
\begin{align}
b_4=\frac{2}{3}(2n_F+n_f)+\frac{1}{3}-\frac{44}{3},\quad
b_L=\frac{4}{3}n_F+\frac{2}{3}-\frac{22}{3},\quad
b_R=\frac{4}{3}n_F-\frac{22}{3}
\end{align}
where $n_F$ and $n_f$ are defined in Table~\ref{tab:matterIR}. As shown in Fig.~\ref{fig:NA4NB2CaseAplots}(a) there are only two viable points with $n_F\geq3$,
\begin{align}
\textrm{P1}:\, n_F=3,\, n_f=15;\quad
\textrm{P2}:\, n_F=4,\, n_f=13,
\label{pp12}\end{align}
that give SAFEs. The corresponding fixed point values of the coupling ratios, for both the stable and unstable cases, are shown in Fig.~\ref{fig:NA4NB2CaseAplots}(b).
\linespread{1}\begin{figure}[!h]
  \centering%
    \includegraphics[width=9cm]{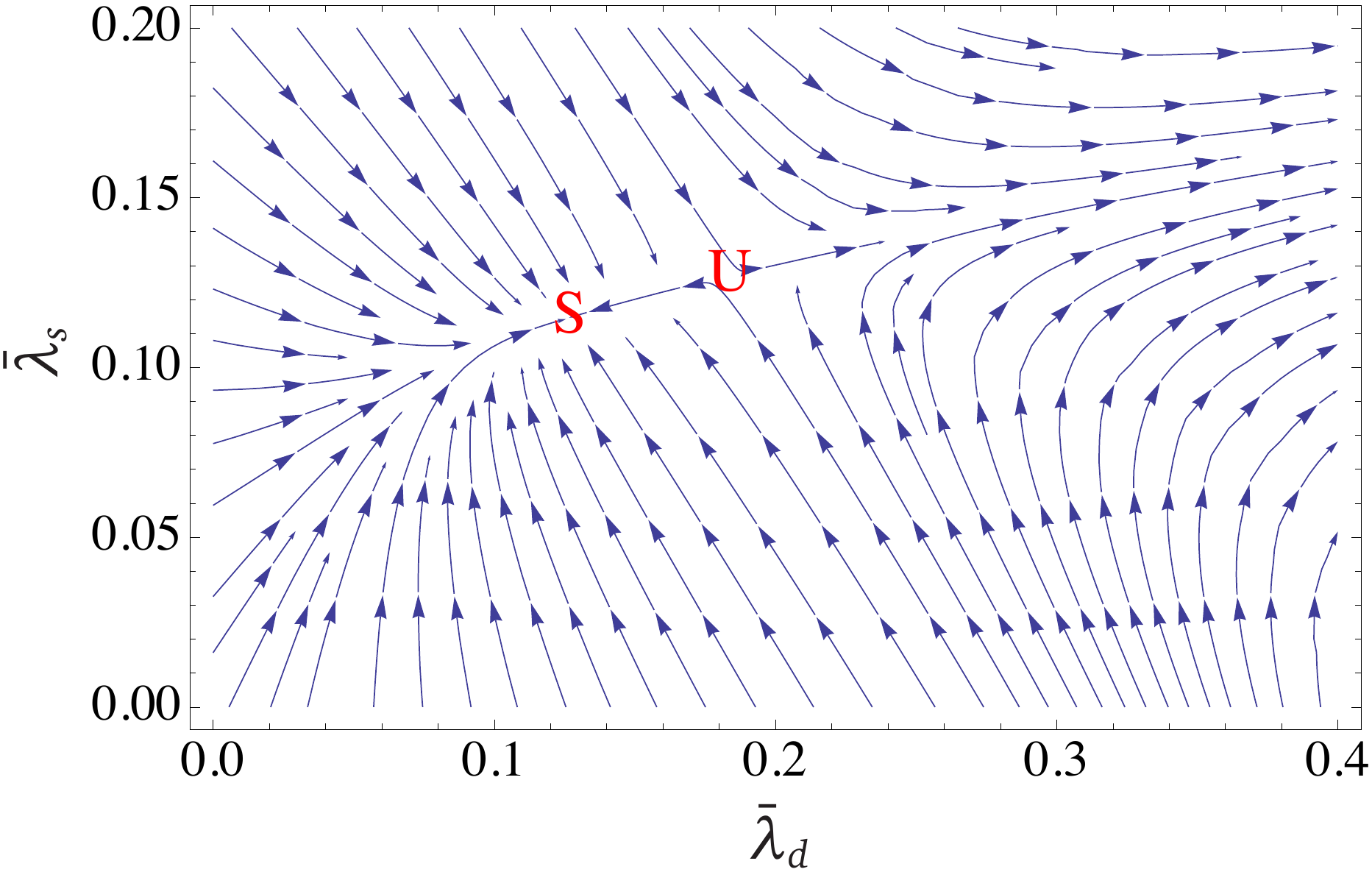}
\caption{\label{fig:flow} Quartic coupling flow towards the UV for the case P1, showing the stable and unstable fixed points. $g_L/g_4$ is set to its fixed point value.}
\end{figure}

Fig.~(\ref{fig:flow}) shows how the quartic couplings flow towards the UV for the case P1. The basin of attraction lies to the left of a line on which the unstable fixed point sits. Although the $SU(2)_L$ gauge coupling $g_L$ is given by its fixed point value  $g_L/g_4=1/8$ for this plot, the basin of attraction hardly changes as long as $g_L/g_4\lesssim 1$ down to some IR scale of interest. For $g_L/g_4\gtrsim 1$ the boundary starts to move significantly to the left, until at $g_L/g_4\approx 2$ the quartic couplings can no longer both be positive at that IR scale. By also imposing the vacuum stability conditions in (\ref{eq:CaseAVS}) on the basin of attraction, we find that the viable flows for the $\bar\lambda_i$ are restricted to a finite region that shrinks if $g_L/g_4$ grows larger.

$SU(4)$ must break at a high enough scale, at least higher than $\mathcal{O}(100)$ TeV, due to constraints for example from the rare decay $K\to e\mu$. (The constraints on $SU(2)_R$ breaking are not so strong.) On the other hand the $(4,2,1)$ scalar $\Phi$ is not available to break $SU(4)$ since the VEV $\langle\Phi\rangle$ would also break $SU(2)_L$. The VEV of an additional $(4,1,2)$ scalar would be sufficient to break the Pati-Salam gauge group down to $SU(3)\times SU(2)_L\times U(1)$, but then the model would not be SAFE. This leaves strong dynamics as the possible origin for this breakdown. We note that the fermion content includes the right-handed neutrino, and a right-handed neutrino condensate does break the Pati-Salam gauge group down in the desired manner.\footnote{The $SU(4)$ preserving condensate $\langle\bar FF\rangle$ would break $SU(2)_L\times SU(2)_R$ but the resistance offered by $SU(2)_L\times SU(2)_R$ to this breaking is enhanced by the number of chiral doublets.} Lepton number is violated, but baryon number and proton stability is preserved.

\begin{figure}[!h]
  \centering%
    \includegraphics[width=6cm]{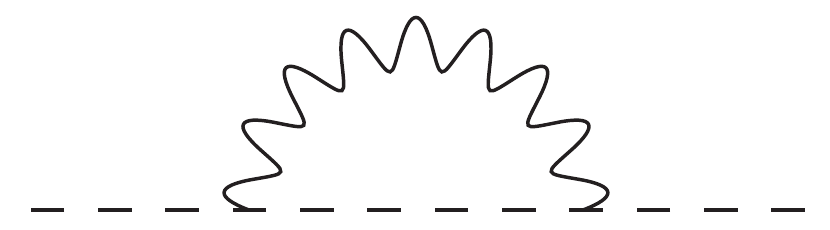}
\caption{\label{fig:gaugehiggs1L} One loop correction to the Higgs mass from an $SU(4)$ gauge boson. }
\end{figure}

Here we see the remaining tension in a low scale unification model because there is still some hierarchy between the unification scale and the Higgs mass that remains unexplained. In our case the neutrino condensate would give rise to a massive $SU(4)$ gauge boson which in turn will contribute to the Higgs mass via the diagram in Fig.~\ref{fig:gaugehiggs1L}. Some other peculiar property of the strong interactions would be needed to explain the suppression of $K\to e\mu$ and the small Higgs mass simultaneously.

\linespread{1}

Another property of the model is that no Yukawa couplings are allowed by the Pati-Salam gauge symmetries. So Yukawa couplings would have to be induced by the same strong interactions that break these symmetries. The resulting couplings are not too constrained by symmetries since they need only respect the unbroken SM symmetries. $SU(2)_R$ is broken and so there is no reason to expect $m_t=m_b$ and $SU(4)$ is broken and so there is no reason to expect $m_b=m_\tau$ etc. Dynamically generated Yukawa couplings may seem peculiar but they just correspond to certain induced three-point amplitudes with soft UV behavior, just as dynamical masses are induced two-point amplitudes with soft UV behavior.

With regard to a strong $SU(4)$ there are two other requirements to meet. The first concerns the impact of higher loop corrections on the $SU(4)$ $\beta$-function. Because the one loop contribution is restricted to be small, the higher order contributions can be relatively large. If these contributions are positive then an infrared fixed point (IRFP) can arise that is approached from below. We need to check that it is large enough for dynamical symmetry breaking. The second requirement is that we need the $SU(3)$ $\beta$-function to turn sufficiently positive below this breaking scale, so that $\alpha_s$ can approach the desired $\sim0.12$ value at the electroweak scale. The fact that it does turn positive is to be expected since $SU(4)$ has a small negative $\beta$-function, and the removal of a negative gauge contribution due to $SU(4)$ breaking can turn it positive. In other words it is the additional vector-like fermions in the model that can produce a positive $SU(3)$ $\beta$-function. Here we find a $SU(3)$ IRFP that is approached from above, but only down to the mass scale of these fermions. These fermion masses could thus be close to a TeV.

From the first requirement the number $n_f'$ of vector-like fermions present in the theory at the $SU(4)$ breaking scale needs to be less than the number $n_f$ listed in (\ref{pp12}). (The fermions not present must have some larger mass.) From the second requirement $n_f'$ cannot be too small. By considering 4-loop $\beta$-functions \cite{vanRitbergen:1997va} we find that perhaps the best compromise is $2n_F+n'_f=15$. Then the $SU(4)$ IRFP is at $\alpha_4\sim0.43$ while the $SU(3)$ IRFP is at $\alpha_s\sim0.12$. The difference between these two numbers is interesting but it is not certain that it is large enough.

\section{Conclusions}
\label{conclusion}

In this paper we explore the construction of UV complete quantum field theories containing truly elementary scalar fields without UV Landau poles. We extend the old study in \cite{Cheng:1973nv} to search for SAFEs for semi-simple gauge groups, which is well motivated to achieve low scale unification. The UV property of gravity is far from clear and we restrict ourselves to study $\beta$-functions of the coupled system of gauge, Yukawa and quartic couplings.

We review the basic idea of a SAFE in Sec.~\ref{review} and present numerical results of simple gauge group for comparison with latter analysis. In Sec.~\ref{general} we generalize the analysis to the semi-simple gauge groups in (\ref{eq:semiLie}), which includes the Pati-Salam model and other low scale unification models as examples. We only consider scalars in fundamental representations, both to incorporate the SM Higgs and to minimize the number of scalar degrees of freedom. We build up four benchmarks for quantitative study and the $\beta$-functions are presented in (\ref{eq:betaNANB}) and Appendix \ref{app:beta}. Our main numerical results and analysis are presented in Sec.~\ref{results}. We search for solutions by parameter scan over gauge group size $N_i$ and $\beta$-function coefficients $b_i$. For each $N_i$ set, we find the upper bounds on $r_i\equiv b_i/b_{i,M}$. To provide a guide for model building, we present these upper bounds in Fig.~\ref{fig:PTbi} and Fig.~\ref{fig:PTbiCaseC2} for all benchmarks. In Sec.~\ref{models} we consider the simplest model that illustrates some of the issues to be faced in SAFE model building.

We list here the properties of the UVFP in SAFEs that we have observed.
\begin{itemize}
\item The gauge couplings and typically most of the quartic couplings are running as $1/t$ in fixed ratios.
\item Stability demands that the Yukawa couplings vanish more rapidly, $1/t^\alpha$ with $\alpha>1$, as do those quartic couplings that have vanishing $\bar\lambda_i$.
\item Fewer scalar degrees of freedom helps to achieve SAFEs.
\item A hierarchy in the sizes of the different gauge groups helps to achieve SAFEs.
\item Among all UVFPs there is always one that is UV stable.
\item SAFEs with negative quartic couplings are rare.
\item The gauge coupling associated with the largest group is typically constrained to run the slowest of all the couplings. Since its associated $b$ is the smallest, it is the largest coupling in the UV.
\item To achieve this small $b$ the theory typically needs some number of vector-like fermions that are only charged under the largest gauge group(s).
\end{itemize}

If the coupling ratios remain anywhere in the vicinity of the fixed point as the couplings themselves grow larger, then it will be the case that the largest gauge group grows strong first in the infrared. This situation may be related to the real world where the quartic couplings and the gauge couplings of the small electroweak gauge groups are observed to be small. In fact in our simplest model we saw that the IR flow of couplings was such that a linear combination of the quartic couplings was bounded from above.

Yukawa $\beta$-functions often have the additional property that they are proportional to the Yukawa couplings to all orders. Thus if they are actually set to their stable fixed point values they are in fact identically vanishing. In our simplest model we saw that the Yukawa couplings identically vanished by gauge symmetry, and thus were only allowed to be generated once the symmetry was broken. It appears that this breakdown and generation of the Yukawa couplings occurs at the scale where the largest gauge group grows strong. The picture is that Yukawa couplings have a dynamical origin in contrast to the truly fundamental gauge and scalar quartic couplings.

It is interesting to compare a SAFE involving several gauge groups to the case of grand unification. In the latter case relations between gauge couplings are fixed by the unification of several gauge groups at some scale. In the SAFE, the ratios of all couplings are flowing to fixed values at the UVFP. But while the theory is fixed in the UV, the theory in the IR is dependent on which flow path the theory is on. A SAFE could be extended to gravity if gravity is also asymptotically free, as is the case for quadratic higher derivative gravity theories. In such a theory all coupling ratios, including gravitational couplings, may be fixed in the deep UV. In this case the ratios of the non gravitational couplings at this ultimate fixed point may differ from what we have described here.

In summary our results show that it might be possible to construct realistic completely asymptotically free gauge theories with complete UV stability containing both fermions and scalars in context of semi-simple gauge groups. This is in contrast to the studies reviewed in \cite{Callaway:1988ya} that typically suffer from UV instability. Our results may be of interest to unification model building beyond the Pati-Salam model, and it can be generalized to incorporate other scalar representations that may be of interest in that context.

\vspace{2ex}\noindent\textbf{Note Added:} As we were finalizing this paper we saw the new paper \cite{Giudice:2014tma}. This paper discusses CAFEs that are not SAFEs, since nonvanishing Yukawa couplings at unstable fixed points are utilized. We also noticed a particular quartic term (third term in their (A.3f)) that we missed that would be present in our Case C with $(4,2,1)$ and $(4,1,2)$ scalars. This term has the same property we discussed for the Levi-Civita term and it does not change the absence of a SAFE in this case. Otherwise our $\beta$-functions agree where they overlap up to the normalization of the quartic couplings.

\begin{acknowledgments}
This research was supported in part by the Natural Sciences and Engineering Research Council of Canada. J.R. is also supported in part by the International Postdoctoral Exchange Fellowship Program of China.
\end{acknowledgments}

\appendix
\section{$\beta$-functions}
\label{app:beta}

In this appendix, we present the one loop $\beta$-functions for the quartic couplings for Cases B, C and D. As explained in Sec.~\ref{general}, Yukawa couplings can be neglected in the scalar $\beta$-functions in the context of SAFEs. Thus our expressions only contain quartic and gauge couplings terms.

For Case B, we deduce the five one loop $\beta$-functions from the potential (\ref{eq:potential2}),
\begin{align}\label{eq:beta2}
(4\pi )^2\beta _{\lambda _{d1}}&=4\left[\left(N_AN_B+4\right)\lambda _{d1}^2+2\left(N_A+N_B\right)\lambda _{d1}\lambda _{s1}+3\lambda _{s1}^2\right]+4\lambda _{d12}\left(N_A\lambda _{d12}+2\lambda _{s12}\right)\nonumber\\
&-6\lambda _{d1}\left[\left(N_A-\frac{1}{N_A}\right)g_A^2+\left(N_B-\frac{1}{N_B}\right)g_B^2\right]\nonumber\\
&+\frac{3}{4}\left[\left(1+\frac{2}{N_A^2}\right)g_A^4+\left(1+\frac{2}{N_B^2}\right)g_B^4\right]+3g_A^2g_B^2\left(1+\frac{1}{N_AN_B}\right)\nonumber\\
(4\pi )^2\beta _{\lambda _{s1}}&=4\lambda _{s1}\left[(N_A+N_B)\lambda _{s1}+6\lambda _{d1}\right]+4\lambda _{s12}^2
-6\lambda _{s1}\left[\left(N_A-\frac{1}{N_A}\right)g_A^2+\left(N_B-\frac{1}{N_B}\right)g_B^2\right]\nonumber\\
&+\frac{3}{4}\left[\left(N_A-\frac{4}{N_A}\right)g_A^4+\left(N_B-\frac{4}{N_B}\right)g_B^4\right]-3g_A^2g_B^2\left(\frac{1}{N_A}+\frac{1}{N_B}\right)\nonumber\\
(4\pi)^2\beta_{\lambda_{2}}&=4\left[(N_A+4)\lambda_{2}^2+N_B\lambda_{d12}(N_A\lambda_{d12}+2\lambda_{s12})+N_B\lambda_{s12}^2\right]
-6\lambda_{2}\left(N_A-\frac{1}{N_A}\right)g_A^2\nonumber\\
&+\frac{3}{4}g_A^4\left[\left(N_A-\frac{4}{N_A}\right)+\left(1+\frac{2}{N_A^2}\right)\right]\nonumber\\
(4\pi )^2\beta _{\lambda _{d12}}&=4\Big[2\lambda _{d12}^2+\lambda _{s12}^2+\lambda _{d1}\left(\left(N_AN_B+1\right)\lambda _{d12}+N_B\lambda _{s12}\right)+\lambda _{s1}\left(\left(N_A+N_B\right)\lambda _{d12}+\lambda _{s12}\right)\nonumber\\
&+\lambda _{2}\left(\left(N_A+1\right)\lambda _{d12}+\lambda _{s12}\right)\Big]\nonumber\\
&-3\lambda _{d12}\left[2\left( N_A-\frac{1}{N_A}\right)g_A^2+\left( N_B-\frac{1}{N_B}\right)g_B^2\right]
+\frac{3}{4}\left(1+\frac{2}{N_A^2}\right)g_A^4\nonumber\\
(4\pi )^2\beta _{\lambda _{s12}}&=4\lambda _{s12}\left(N_A\lambda _{s12}+4\lambda _{d12}+N_B\lambda _{s1}+\lambda _{d1}+\lambda _{2}\right)\nonumber\\
&-3\lambda _{s12}\left[2\left( N_A-\frac{1}{N_A}\right)g_A^2+\left( N_B-\frac{1}{N_B}\right)g_B^2\right]
+\frac{3}{4}\left(N_A-\frac{4}{N_A}\right)g_A^4
\end{align}
where $N_A$ denotes the common gauge group.

Case C is split into two benchmarks. In Case C1, by imposing $Z_2$ symmetry as in (\ref{eq:potential3}), we deduce one loop $\beta$-functions for the four quartic couplings from (\ref{eq:caseC1Z2}).
\begin{align}\label{eq:beta31}
(4\pi )^2\beta _{\lambda _{d1}}&=4\left[\left(N_AN_B+4\right)\lambda _{d1}^2+2\left(N_A+N_B\right)\lambda _{d1}\lambda _{s1}+3\lambda _{s1}^2\right]+4N_B\lambda _{d12}\left(N_A\lambda _{d12}+2\lambda _{s12}\right)\nonumber\\
&-6\lambda _{d1}\left[\left(N_A-\frac{1}{N_A}\right)g_A^2+\left(N_B-\frac{1}{N_B}\right)g_B^2\right]\nonumber\\
&+\frac{3}{4}\left[\left(1+\frac{2}{N_A^2}\right)g_A^4+\left(1+\frac{2}{N_B^2}\right)g_B^4\right]+3g_A^2g_B^2\left(1+\frac{1}{N_AN_B}\right)\nonumber\\
(4\pi )^2\beta _{\lambda _{s1}}&=4\lambda _{s1}\left[(N_A+N_B)\lambda _{s1}+6\lambda _{d1}\right]+4N_B\lambda _{s12}^2
-6\lambda _{s1}\left[\left(N_A-\frac{1}{N_A}\right)g_A^2+\left(N_B-\frac{1}{N_B}\right)g_B^2\right]\nonumber\\
&+\frac{3}{4}\left[\left(N_A-\frac{4}{N_A}\right)g_A^4+\left(N_B-\frac{4}{N_B}\right)g_B^4\right]-3g_A^2g_B^2\left(\frac{1}{N_A}+\frac{1}{N_B}\right)\nonumber\\
(4\pi )^2\beta _{\lambda _{d12}}&=4\Big[2\lambda _{d12}^2+\lambda _{s12}^2+2\lambda _{d1}\left(\left(N_AN_B+1\right)\lambda _{d12}+N_B\lambda _{s12}\right)+2\lambda _{s1}\left(\left(N_A+N_B\right)\lambda _{d12}+\lambda _{s12}\right)\nonumber\\
&-6\lambda _{d12}\left[\left( N_A-\frac{1}{N_A}\right)g_A^2+\left( N_B-\frac{1}{N_B}\right)g_B^2\right]+\frac{3}{4}\left(1+\frac{2}{N_A^2}\right)g_A^4\nonumber\\
(4\pi )^2\beta _{\lambda _{s12}}&=4\lambda _{s12}\left(N_A\lambda _{s12}+4\lambda _{d12}+2N_B\lambda _{s1}+2\lambda _{d1}\right)\nonumber\\
&-6\lambda _{s12}\left[\left( N_A-\frac{1}{N_A}\right)g_A^2+\left( N_B-\frac{1}{N_B}\right)g_B^2\right]
+\frac{3}{4}\left(N_A-\frac{4}{N_A}\right)g_A^4
\end{align}

Case C2 denotes the general case without $Z_2$ symmetry. The one loop $\beta$-functions for the six quartic couplings are
\begin{align}\label{eq:beta32}
(4\pi )^2\beta _{\lambda _{d1}}&=4\left[\left(N_AN_B+4\right)\lambda _{d1}^2+2\left(N_A+N_B\right)\lambda _{d1}\lambda _{s1}+3\lambda _{s1}^2\right]+4N_C\lambda _{d12}\left(N_A\lambda _{d12}+2\lambda _{s12}\right)\nonumber\\
&-6\lambda _{d1}\left[\left(N_A-\frac{1}{N_A}\right)g_A^2+\left(N_B-\frac{1}{N_B}\right)g_B^2\right]\nonumber\\
&+\frac{3}{4}\left[\left(1+\frac{2}{N_A^2}\right)g_A^4+\left(1+\frac{2}{N_B^2}\right)g_B^4\right]+3g_A^2g_B^2\left(1+\frac{1}{N_AN_B}\right)\nonumber\\
(4\pi )^2\beta _{\lambda _{s1}}&=4\lambda _{s1}\left[(N_A+N_B)\lambda _{s1}+6\lambda _{d1}\right]+4N_C\lambda _{s12}^2
-6\lambda _{s1}\left[\left(N_A-\frac{1}{N_A}\right)g_A^2+\left(N_B-\frac{1}{N_B}\right)g_B^2\right]\nonumber\\
&+\frac{3}{4}\left[\left(N_A-\frac{4}{N_A}\right)g_A^4+\left(N_B-\frac{4}{N_B}\right)g_B^4\right]-3g_A^2g_B^2\left(\frac{1}{N_A}+\frac{1}{N_B}\right)\nonumber\\
(4\pi )^2\beta _{\lambda _{d2}}&=(4\pi )^2\beta _{\lambda _{d1}}(N_B\to N_C, \,g_B\to g_C, \,\lambda_{d1}\to\lambda_{d2},\, \lambda_{s1}\to\lambda_{s2})\nonumber\\
(4\pi )^2\beta _{\lambda _{s2}}&=(4\pi )^2\beta _{\lambda _{s1}}(N_B\to N_C, \,g_B\to g_C, \,\lambda_{d1}\to\lambda_{d2},\, \lambda_{s1}\to\lambda_{s2})\nonumber\\
(4\pi )^2\beta _{\lambda _{d12}}&=4\Big[2\lambda _{d12}^2+\lambda _{s12}^2+\lambda _{d1}\left(\left(N_AN_B+1\right)\lambda _{d12}+N_B\lambda _{s12}\right)+\lambda _{s1}\left(\left(N_A+N_B\right)\lambda _{d12}+\lambda _{s12}\right)\nonumber\\
&+\lambda _{d2}\left(\left(N_AN_C+1\right)\lambda _{d12}+N_C\lambda _{s12}\right)+\lambda _{s2}\left(\left(N_A+N_C\right)\lambda _{d12}+\lambda _{s12}\right)\Big]\nonumber\\
&-3\lambda _{d12}\left[2\left( N_A-\frac{1}{N_A}\right)g_A^2+\left( N_B-\frac{1}{N_B}\right)g_B^2+\left( N_C-\frac{1}{N_C}\right)g_C^2\right]\nonumber\\
&+\frac{3}{4}\left(1+\frac{2}{N_A^2}\right)g_A^4\nonumber\\
(4\pi )^2\beta _{\lambda _{s12}}&=4\lambda _{s12}\left(N_A\lambda _{s12}+4\lambda _{d12}+N_B\lambda _{s1}+N_C\lambda _{s2}+\lambda _{d1}+\lambda _{d2}\right)-3\lambda _{s12}\left[2\left( N_A-\frac{1}{N_A}\right)g_A^2\right.\nonumber\\
&\left.+\left( N_B-\frac{1}{N_B}\right)g_B^2+\left( N_C-\frac{1}{N_C}\right)g_C^2\right]
+\frac{3}{4}\left(N_A-\frac{4}{N_A}\right)g_A^4
\end{align}
They are symmetric under interchange $\lambda_{d1}\rightarrow\lambda_{d2}$, $\lambda_{s1}\rightarrow\lambda_{s2}$, $N_B\rightarrow N_C$ and $g_B\rightarrow g_C$. When $N_A=4, N_B=N_C=2$, a new quartic coupling can be constructed by the Levi-Civita symbol as in (\ref{eq:potential31LC}). The $\beta$-functions are then modified as
\begin{align}\label{eq:beta32LC1}
(4\pi)^2\beta_{di}\to(4\pi)^2\beta_{di}+8\lambda_\epsilon^2,\quad
(4\pi)^2\beta_{si}\to(4\pi)^2\beta_{si}-8\lambda_\epsilon^2.
\end{align}
The $\beta$-function of this new coupling is
\begin{align}\label{eq:beta32LC2}
(4\pi)^2\beta_{\epsilon}&=4\lambda_\epsilon\left[\lambda_{d1}+\lambda_{d2}-\lambda_{s1}-\lambda_{s2}+4(\lambda_{d12}-\lambda_{s12})\right]-\frac{9}{2}\lambda_\epsilon (5g_4^2+g_L^2+g_R^2).
\end{align}

For Case D, there are one double trace and three single trace couplings. From potential (\ref{eq:potentialsNANBNC}), we deduce following $\beta$-functions,
\begin{align}\label{eq:betaNANBNC}
(4\pi )^2\beta _{\lambda _{d}}&=4 \Big[\lambda _d^2 \left(N_A N_B N_C+4\right)+2 \lambda _d \left(\lambda _{s1} \left(N_A+N_B N_C\right)+\lambda _{s2} \left(N_A N_C+N_B\right)+\lambda _{s3} \left(N_A N_B+N_C\right)\right)\nonumber\\
&+2 N_A \lambda _{s2} \lambda _{s3}+2 N_B \lambda _{s1} \lambda _{s3}+2 N_C \lambda _{s1} \lambda _{s2}+3 \left(\lambda _{s1}^2+\lambda _{s2}^2+\lambda _{s3}^2\right)\Big]\nonumber\\
&-6\lambda _{d}\left[\left(N_A-\frac{1}{N_A}\right)g_A^2+\left(N_B-\frac{1}{N_B}\right)g_B^2+\left(N_C-\frac{1}{N_C}\right)g_C^2\right]\nonumber\\
&+\frac{3}{4}\left[\left(1+\frac{2}{N_A^2}\right)g_A^4+\left(1+\frac{2}{N_B^2}\right)g_B^4+\left(1+\frac{2}{N_C^2}\right)g_C^4\right]
+3\left(\frac{g_A^2g_B^2}{N_AN_B}+\frac{g_A^2g_C^2}{N_AN_C}+\frac{g_B^2g_C^2}{N_BN_C}\right)\nonumber\\
(4\pi )^2\beta _{\lambda _{s1}}&=4 \Big[\lambda _{s1}^2 \left(N_B N_C+N_A\right)+2 \lambda _{s1} \left(3 \lambda _d+N_B \lambda _{s2}+N_C \lambda _{s3 }\right)+4 \lambda _{s2} \lambda _{s3}\Big]\nonumber\\
&-6\lambda _{s1}\left[\left(N_A-\frac{1}{N_A}\right)g_A^2+\left(N_B-\frac{1}{N_B}\right)g_B^2+\left(N_C-\frac{1}{N_C}\right)g_C^2\right]\nonumber\\
&+\frac{3}{4}g_A^4\left(N_A-\frac{4}{N_A}\right)+3\left[g_B^2 g_C^2-g_A^2 \left(\frac{g_B^2}{N_B}+\frac{g_C^2}{N_C}\right)\right]\nonumber\\
(4\pi )^2\beta _{\lambda _{s2}}&=4 \Big[\lambda _{s2}^2 \left(N_A N_C+N_B\right)+2 \lambda _{s2} \left(3 \lambda _d+N_A \lambda _{s1}+N_C \lambda _{s3 }\right)+4 \lambda _{s1} \lambda _{s3}\Big]\nonumber\\
&-6\lambda _{s2}\left[\left(N_A-\frac{1}{N_A}\right)g_A^2+\left(N_B-\frac{1}{N_B}\right)g_B^2+\left(N_C-\frac{1}{N_C}\right)g_C^2\right]\nonumber\\
&+\frac{3}{4}g_B^4\left(N_B-\frac{4}{N_B}\right)+3\left[g_A^2 g_C^2-g_B^2 \left(\frac{g_A^2}{N_A}+\frac{g_C^2}{N_C}\right)\right]\nonumber\\
(4\pi )^2\beta _{\lambda _{s3}}&=4 \Big[\lambda _{s3}^2 \left(N_A N_B+N_C\right)+2 \lambda _{s3} \left(3 \lambda _d+N_A \lambda _{s1}+N_B \lambda _{s2 }\right)+4 \lambda _{s1} \lambda _{s2}\Big]\nonumber\\
&-6\lambda _{s3}\left[\left(N_A-\frac{1}{N_A}\right)g_A^2+\left(N_B-\frac{1}{N_B}\right)g_B^2+\left(N_C-\frac{1}{N_C}\right)g_C^2\right]\nonumber\\
&+\frac{3}{4}g_C^4\left(N_C-\frac{4}{N_C}\right)+3\left[g_A^2 g_B^2-g_C^2 \left(\frac{g_A^2}{N_A}+\frac{g_B^2}{N_B}\right)\right]
\end{align}
For $N_A=4, N_B=N_C=2$ case, the modification of $\beta$-functions from the Levi-Civita term in (\ref{eq:CaseDLC}) is quite similar to that in Case C. We find
\begin{align}\label{eq:betaNANBNCLC1}
(4\pi)^2\beta_{d}\to(4\pi)^2\beta_{d}+8\lambda_\epsilon^2,\quad
(4\pi)^2\beta_{s1}\to(4\pi)^2\beta_{s1}-8\lambda_\epsilon^2.
\end{align}
The $\beta$-function of this new coupling is
\begin{align}\label{eq:betaNANBNCLC2}
(4\pi)^2\beta_{\epsilon}&=24\lambda_\epsilon\left(\lambda_{d}-\lambda_{s1}\right)-\frac{9}{2}\lambda_\epsilon (5g_4^2+2g_L^2+2g_R^2).
\end{align}

\linespread{1}

\end{document}